\documentclass{article}
\usepackage{amsmath,amssymb,geometry}
\usepackage{sfmath}
\usepackage{helvet}
\usepackage{mathtools}
\usepackage{comment}
\usepackage{tgtermes}
\usepackage{cmbright}

\usepackage{natbib}	
\usepackage{bibentry}
\usepackage{setspace}
\usepackage{hyperref}
\author{Vikram Manjunath\thanks{\raggedright University of Ottawa; \texttt{vikram@dosamobile.com}.}  \and Alexander
Westkamp\thanks{\raggedright Department of Management, Economics and
  Social Sciences, University of Cologne;
  \texttt{westkamp@wiso.uni-koeln.de}}}

\usepackage{amsthm}
\newtheorem{proposition}{Proposition}
\newtheorem{theorem}{Theorem}
\newtheorem{corollary}{Corollary}
\newtheorem{example}{Example}

\newtheorem{lemma}{Lemma}
\newtheorem{claim}{Claim}

\usepackage[capitalize]{cleveref}
\Crefname{example}{Example}{Examples}
\Crefname{figure}{Figure}{Figures}
\Crefname{theorem}{Theorem}{Theorems}
\Crefname{section}{Section}{Sections}

\usepackage{graphicx}
\usepackage{subfig}

\hypersetup{colorlinks=true,linkcolor=black,citecolor=black,urlcolor=blue}

\usepackage{xspace}

\newif\ifdraft

\newcommand{\aw}[1]{{\ifdraft\color{blue}\fi #1}\xspace}

\newcommand{\A}{attractive\xspace}
\newcommand{\B}{bearable\xspace}
\newcommand{\C}{crummy\xspace}

\title{Marginal Mechanisms for Balanced Exchange}

\begin{document}
\maketitle

\begin{abstract}
We study balanced exchange problems in which agents with responsive preferences are initially endowed with multiple indivisible objects and can trade without transfers, as in shift exchange or time-banking. In many such settings, eliciting or processing full preferences over bundles is infeasible. Instead, mechanisms rely solely on \emph{marginal preferences}, that is, rankings of individual objects. 

We characterize when eliciting only marginal preferences is enough to \emph{unambiguously} identify allocations that are  \emph{efficient} and \emph{individually rational} in the sense that 
these properties hold  with respect to \underline{any} responsive preferences consistent with the elicited marginals. We parameterize domains of marginal preferences  by which indifference classes can contain endowed and non-endowed objects. We show that the essentially unique maximal domain for which an unambiguously efficient and unambiguously individually rational marginal mechanism exists is \emph{trichotomous}:  Agents rank objects in three tiers, with the bottom tier not containing any endowed objects. 

We also consider agents' incentives for truthful preference revelation. The maximal domain for which an efficient, individually rational, and strategy-proof mechanism, marginal or not, exists is \emph{strongly trichotomous}: Agents rank objects in three tiers, with the bottom tier not containing any endowed objects and the middle tier not containing any non-endowed objects. The canonical marginal mechanism that unambiguously achieves our three desiderata on that domain is a \emph{serial dictatorship} over the set of individually rational allocations. Interestingly, when employed on the larger trichotomous domain, this serial dictatorship mechanism still has a weakly dominant strategy: Reveal the top preference tier truthfully and do not reveal any non-endowed objects in the middle  tier. We propose a new family of \emph{gradual-revelation mechanisms} that are also unambiguously efficient and unambiguously individually rational on the trichotomous domain, while providing ``better'' incentives for the truthful revelation of all three preference tiers.
\end{abstract}


\bigskip

\newpage

\section{Introduction}

An interesting class of exchange problems involves agents, who are endowed with multiple indivisible objects and can trade some or all of their endowed objects, but  can neither use monetary transfers nor change the total number objects they are assigned. Examples of such ``balanced'' reallocation problems include the exchange of work shifts when the total number of shifts per worker is fixed \citep{ManjunathWestkamp:2018a} or the exchange of service hours in time banks when each user's number of demanded hours is required to equal her number of supplied hours \citep{AnderssonCsehEhlersErlanson:2018}. Such environments are challenging since each participant ultimately consumes a \emph{bundle} of objects,  yet in practice it is  unrealistic to process or even elicit full preferences over all feasible bundles given exponential growth in the number of individual objects. This point has been made  in work on practical market design and preference elicitation \citep{Milgrom:AEJmicro2009,Milgrom:GEB2010,Milgrom:EI2011,OthmanSandholmBudish:AAMAS2010,BudishKessler:2018}. A more realistic approach is to elicit only \emph{marginal} information about preferences by only asking agents to rank objects. Such reporting languages are commonplace in real-world exchange platforms in applications such as scheduling or shift exchange.\footnote{These platforms also often only consider the possibility of bilateral swaps, thus further reducing efficiency of the exchange process. For example, see the following  for a description of how shift exchange is implemented in Microsoft Teams: \url{https://support.microsoft.com/en-us/office/request-open-shifts-swap-or-offer-shifts-in-shifts-50920ad2-a0e7-4f4a-aeb8-adf57d3b0c88}.} We study what can and cannot be achieved when a designer restricts attention to mechanisms that take only this marginal information as input.

A useful starting point is the theoretical analysis of many-to-one matching, say between workers (who each want at most one job) and firms (who seek to fill  multiple positions). Firms' preferences over groups are \emph{responsive} to marginal preferences over individual workers. \citet{Roth1985}  shows that these marginal preferences are enough to find and verify matchings that are stable no matter how the marginals extend to preferences over groups. Moreover, even though the marginal preference of a firm does not pin down how it ranks two arbitrary sets of workers, it does pin down how the firm ranks sets of workers to whom it is matched by  stable matchings. In our  balanced exchange setting, the natural allocative desiderata are efficiency and individual rationality. Here, an analogous ``marginals suffice'' intuition fails: marginal rankings generally do \emph{not} determine which reallocations are efficient and individually rational without any ambiguity. That is, it can be that no reallocation is efficient and individually rational \emph{for all} responsive preferences consistent with a given profile of marginal rankings. Thus, while responsiveness is enough to make stability tractable in many-to-one matching, it does not eliminate the fundamental ambiguity of bundle comparisons that arises in multi-unit exchange without transfers. In this paper we ask when marginal reports can support strong welfare and participation guarantees despite the this ambiguity, and when they can also support meaningful incentive properties?

To motivate our idea of how to define domains of marginal preferences, consider a real-life example of \emph{shift exchange} in the medical sector. Each member of a team of physicians is initially assigned a set of shifts for a scheduling period (for example, a "call schedule"). They can trade shifts, but each must still cover the same number of shifts overall, so trading has to be balanced. Real systems implement such exchanges through simple user interfaces that often give users only two or three tiers to rank shifts (e.g. the ``yes,'' ``no,'' and ``maybe/if-need-be'' options that are often available in scheduling tools). These interfaces allow physicians to express the presence or absence of soft or hard scheduling constraints that may limit their availabilities. Outside of the specific context of shift exchange, the design choice to provide agents with only a few tiers to rank objects seems especially natural in everyday and low-stakes settings. Here, agents may be unwilling to invest effort in providing detailed information to the mechanism designer, and may also be less inclined to deliberate over nuanced rankings.\footnote{While the welfare gains from any single instance may be modest when stakes are low, small performance increases accumulate when these markets operate repeatedly. Hence, even in low-stakes everyday environments, it is worthwhile to focus on how to design  efficient mechanisms.} Furthermore, evidence from marketing suggests that consumers tend to exhibit weaker and less discriminating preferences when faced with larger choice sets \citep{GreenleafLehmann:JCR1995,Dhar:JCR1997,IyengarLepper:2000JPSP,Chernev:JCR2003}. Finally, even in surveys, coarse response scales can be the most reliable measurement instruments \citep{Alwin2007}. Our main results show that if agents' true marginal preferences have no more than three tiers, then we can use these marginals to find reallocations that are \emph{unambiguously} efficient and unambiguously individually rational in that they satisfy these two properties for any responsive preferences consistent with the elicited marginals. Furthermore, in a sense to be made more precise below, there is no larger domain of marginal preferences for which a mechanism that only relies on these marginals is unambiguously efficient and unambiguously individually rational. These results provide a foundation for the use of coarse preference reporting tools that we see in many real-life applications. 

Abstractly, our model involves agents  who are endowed with bundles of indivisible objects and can exchange some or all of their assigned objects without using monetary transfers. Exchange is necessarily  balanced in the sense that each agent has to end up with the same number of objects as she was initially endowed with. Agents' preferences over bundles are responsive to their marginal preferences over individual objects. As explained above, we focus on exchange mechanisms that only elicit these marginal preferences and ask whether we can achieve  efficiency and individually rationality unambiguously. In order to characterize when this goal can be achieved, we consider domains of marginal preferences that are described by a limited menu of tiers that agents can use to rank objects,  with restrictions on which tiers can contain endowed and which tiers can contain non-endowed objects. Such domains capture a wide range,  from trivial examples (e.g. the case where no information is elicited) to giving agents full flexibility to express any marginal preference (the number of tiers coincides with the number of objects and agents do not face any restrictions on which tiers may contain which items). Importantly, we think of a domain (of marginal preferences) as a tool that allows agents to report their true preferences and then ask whether we can unambiguously achieve efficiency and individual rationality on that domain. We do not consider the different, but also interesting, question of what happens when agents with unrestricted true preferences face a pure communication constraint imposed by the mechanism designer. 

Our first main result, mentioned above, identifies a natural domain for which marginal mechanisms can be unambiguously efficient and unambiguously individually rational: Agents  partition objects into attractive (top tier), bearable (middle tier), and  crummy (bottom tier) sets subject to the restriction that no endowed objects can be in the bottom tier. We refer to this as the \emph{trichotomous domain} and show how the two key desiderata can be achieved via an \emph{component-wise individually rational priority (CIRP)} mechanism. Our CIRP mechanism uses agents' reported marginal preferences to first compute the set of all unambiguously individually rational reallocations and then runs a serial dictatorship in which agents take turns  picking their preferred reallocations. Our second main result shows that under mild richness conditions, the trichotomous domain is the essentially unique maximal domain of marginal preferences for which a marginal mechanism can unambiguously achieve efficiency and individual rationality.\footnote{One could consider agents who express arbitrary preferences over non-endowed objects that they rank below all of their endowed objects. Given that unambiguous individual rationality requires agents to never receive any objects from below their second preference tier, the added generality is immaterial.} 

We then consider agents' incentives to reveal their marginal preferences truthfully.  \citet{ManjunathWestkamp:2018a} show that  CIRP mechanisms are \emph{strategy-proof} on the subdomain of \emph{strongly trichotomous preferences}, where agents never rank non-endowed objects in their middle preference tier. We complement this result by showing that the strongly trichotomous domain is the essentially unique maximal domain for which any mechanism, marginal or not, can simultaneously be efficient, individually rational, and strategy-proof.  

Finally, we consider agents' incentives on the trichotomous domain, where full strategy-proofness cannot be achieved but marginal mechanisms can still unambiguously achieve efficiency and individual rationality. Interestingly,  CIRP mechanisms still have a weakly dominant strategy on this larger domain: Each agent is always best-advised to reveal her top tier truthfully and to \emph{hide} all non-endowed bearable objects. The proof is technically involved and establishing it requires new structural arguments about how priority rules behave when a single object is moved across categories in a trichotomous report. Motivated by this negative finding about the incentives of the simple CIRP mechanisms on the trichotomous domain, we propose a new family of unambiguously efficient and unambiguously individually rational mechanisms that avoid the most pathological incentive features of CIRP. These \emph{gradual revelation} (GR) mechanisms delay using information about the middle tier of an agent's marginal preference until  it is guaranteed that the agent can no longer improve without making other agents worse off. A key foundation for the construction of our GR mechanisms is that when agents rank more non-endowed objects in their middle preference tiers, there might be further efficiency enhancing exchanges but it is never possible that the welfare of all agents strictly improves. Hence, there is always at least one agent, whose welfare cannot be improved any further no matter which non-endowed objects are in other agents' middle indifference classes. We can use information about which non-endowed objects are in such an agent's middle preference tier without adversely affecting her incentives to reveal these objects. As a result, truthful reporting is not weakly dominated as it is under the CIRP benchmark. Furthermore, GR mechanisms satisfy two economically motivated incentive properties: truncation-proofness (agents do not gain by dropping acceptable options) and non-obvious manipulability in the sense of \cite{troyanmorrill}. This gives us a sharper picture of the trade-off between robustness and incentives in marginal mechanism design for balanced exchange, together with mechanisms that better match the informational and strategic constraints of practical exchange platforms.

\paragraph{Relation to the literature.}
A substantial literature, beginning with \citet{ShapleyScarf:JME1974}, studies exchange in the benchmark setting where each agent is endowed with and consumes a single object. Under strict preferences, the core is single-valued \citep{RothPostlewaite:JME1977} and induces a strategy-proof mechanism \citep{Roth:EL1982}. Even in the presence of ties, strategy-proofness, efficiency, and individual rationality remain compatible \citep{AlcaldeMolis:GEB2011,JaramilloManjunath:JET2012}. These results do not extend to multi-unit generalizations of the model \citep{Sonmez:1999ecta,KonishiQuintWako:JME2001,BiroKlijnPapai:2018}. Analyses of multi-unit exchange have therefore focused on weakening some desiderata or restricting preferences \citep{PapaiJET2001,Papai:JME2003,EhlersKlausSCW2003,TodoSunYokoo2014,BiroKlijnPapai:2018}. Our contribution is to delineate, within a broad and practically motivated family of marginal domains, exactly when robust welfare and participation guarantees are compatible, when dominant-strategy incentives can be added, and what incentive failures arise at the boundary between these domains. We then introduce new mechanisms and technical tools to analyze manipulability in the trichotomous environment.

Our “unambiguous” welfare notions are robustness requirements under partial preference information. Unambiguous efficiency has been studied in  fair-division and computational social choice (see, for example, \citet{BramsEdelmanFishburnTD2003,BouveretEndrissLang2010,AzizEtAl2019}). However, unambiguous individual rationality appears to be new in our setting.

Finally, this work also contributes to the literature on  coarse preference models, including  dichotomous preferences (see, for example \citet{BogomolnaiaMoulin:Econometrica2004,Ortega2020,aziz2025strategyproofmaximummatchingdichotomous}). Our results  help understand  exactly why such domains are tractable.

\paragraph{Organization.}
\cref{sec:model} introduces the balanced exchange model, responsiveness, and unambiguous welfare  criteria. \cref{sec:compatibility IR E} establish the compatibility and maximality results for unambiguous efficiency and unambiguous individual rationality. \cref{sec: sp max} turns to strategy-proofness and proves the strongly trichotomous maximality. \cref{sec:incentives trichotomous} analyzes strategic behavior on the trichotomous domain and proves the dominant-strategy manipulation result for the priority benchmark. It also defines the gradual revelation mechanisms and establishes their welfare and incentive  properties.

\section{Model}\label{sec:model}

Let $I=\{1,\ldots,n\}$ be a set of $n$ agents and $O$ be a finite set of objects.
Each agent $i\in I$ is endowed with a non-empty set of objects
$\omega(i)\subseteq O$ such 
that $\omega(i) \cap 
\omega(j) = \emptyset$ whenever $i\neq j$ and  $O=\cup_{i\in
  i}\omega(i)$. An \emph{allocation} is a mapping $\mu:I\rightarrow 2^O$ that satisfies the following two requirements:
\begin{enumerate}
\item For any pair of distinct agents $i,j\in I$, $\mu(i)\cap \mu(j)=\emptyset$.
\item For any agent $i\in I$, $|\mu(i)|=|\omega(i)|$.
\end{enumerate}
Thus, the endowment $\omega$, is itself an allocation.
Let $\mathcal M$ be the set of all allocations.
Our definition of an allocation requires the exchange of objects to be
\emph{balanced} in the sense that each agent  ends up with the same number
of objects as she was endowed with. We assume throughout that the endowment is fixed and known to the mechanism designer.\footnote{In our key motivating application to shift exchange, the assumption of an initial shift schedule that is known to the designer is sensible when the designer represents the organization at which participants of the platform are employed.} Due to balancedness, the set of bundles that an agent $i\in I$ may consume at some allocation
in $\mathcal M$, her \emph{consumption set}, is $\mathcal X_i = \{
  X_i \subseteq O : |X_i| = |\omega(i)|
  \}$.

Each agent $i\in I$ has a  preference relation
$R_i$ over $\mathcal X_i$, whose asymmetric part is $P_i$.\footnote{In other words, $R_i$ is a
  complete, transitive, and reflexive relation over $\mathcal X_i$.}
Let $\mathcal R_i$ be the set of permissible preference relations for $i$. Let
$\mathcal R = \times_{i\in I} \mathcal R_i$ be the set of permissible 
preference profiles. Given a preference profile  $R = (R_i)_{i\in I}\in \mathcal R$ and an
allocation $\mu\in \mathcal M$, $\mu$ is  \emph{individually rational
  under $R$} if each agent finds her assignment to be at least as
good as her  endowment: for each $i\in I$, $\mu(i) \mathrel R_i
\omega(i)$. Note that there is always at least one individually
rational allocation: the endowment itself. Next,  $\mu'\in \mathcal M$
\emph{welfare-improves}  upon $\mu\in \mathcal M$ under $R$ if each
agent likes $\mu'$ at least as well as $\mu$ and some agent likes
$\mu'$ strictly better than $\mu$: for each $i\in I$,
$\mu'(i) \mathrel R_i \mu(i)$ and, for some $i\in I$, $\mu'(i)\mathrel P_i \mu(i)$. If
there is no $\mu'$ that welfare-improves upon $\mu$, then $\mu$ is
\emph{efficient} at $R$. We say that  $\mu'\in \mathcal M$
\emph{weakly welfare-improves}  upon $\mu\in \mathcal M$ under $R$ if
each agent likes $\mu'$ at least as well as $\mu$: for each
$i\in I$, $\mu'(i) \mathrel R_i \mu(i)$.  We say that  $\mu'\in \mathcal M$
\emph{strictly welfare-improves}  upon $\mu\in \mathcal M$ under $R$ if
each agent likes $\mu'$ better than $\mu$: for each
$i\in I$, $\mu'(i) \mathrel P_i \mu(i)$. If no other allocation strictly
welfare-improves $\mu$, then we say that $\mu$ is \emph{weakly efficient}.

We call the set of preference profiles   a \emph{domain}. By definition, a domain $\mathcal R$  is the Cartesian product of a set of permissible preferences for each agent. Given a domain $\mathcal R$,
a \emph{mechanism on $\mathcal R$} is a mapping $\varphi:\mathcal R\to
\mathcal M$  that associates each profile in $\mathcal R$ with an allocation. Given a profile $R\in\mathcal R$, we denote the set of objects
that $i$ receives under mechanism $\varphi$ by $\varphi_i(R)$. A
mechanism is \emph{individually rational} if it always (for any
profile of preferences in its domain) selects an individually rational
allocation, and \emph{efficient} if it always selects a
efficient allocation. Finally, a mechanism $\varphi$ is
\emph{strategy-proof} on $\mathcal R$ if no agent can ever benefit by reporting a
preference that is different from her true one: there do not exist
$R\in\mathcal  R$, $i\in I$, and $ {R}_i'\in \mathcal R_i$
such that $\varphi_i(R_i', R_{-i}) \mathrel P_i 
\varphi_i(R)$.

\subsection{Responsive preferences and marginal mechanisms}

Agent $i$'s preference relation $R_i$ over
bundles in $\mathcal X_i$ is \emph{responsive} if there is a weak ordering $\succsim_i$
over $O$ such that,  for any pair of objects 
$o,p\in O$ and any set $Q\in \mathcal X_i$ where $o\in Q$,
$Q\mathrel R_i(Q\setminus \{o\})\cup\{p\}$ if and only if $o\succsim_ip$. We say
that $R_i$ is \emph{responsive to}  $\succsim_i$  and refer to the
latter as $i$'s \emph{marginal preference}. We assume throughout the
remainder of this paper that
each agent has a responsive preference over bundles and, therefore,
that her marginal preferences are well defined. Note that any responsive
preference relation $R_i$ is responsive to a unique marginal
preference relation $\succsim_i$. However, there may be several
preference relations that are all responsive to the same
marginal preference relation. Let $\mathcal{S}_i$ denote the set 
of all marginal preferences  for $i$ and $\mathcal{S}\equiv
\times_{i\in I}\mathcal{S}_i$ denote the set of all marginal
preference profiles.  

It is well understood  (and easy to see) that even for responsive
preferences, marginal preferences do not fully describe agents'
preferences over bundles. However, it is  difficult to rely
on additional information about agents' preferences since (a) it is typically not possible to
 elicit agents' preferences over bundles in an incentive
compatible way and (b) agents may find it significantly more difficult
to describe their full preferences over bundles. It is therefore worth
focusing on \emph{marginal mechanisms} that only rely on
agents' marginal preferences. Formally,  a mechanism
$\varphi$ is \emph{marginal} if for marginal preference profiles $\succsim\in\mathcal{S}$ and any pair of preference profiles $R$ and $R'\in \mathcal R$ that are both responsive to $\succsim$, $\varphi(R)=\varphi(R')$. For marginal mechanisms, we can
think of agents as submitting just their marginal 
preferences. In this sense, if $\varphi$ is marginal, it can be expressed as a function from $\mathcal S$ to $\mathcal M$. Strategy-proofness then simply requires that for each
agent, submitting her marginal preferences truthfully is a weakly
dominant strategy no matter how she extends her marginal preference to
bundles.

\subsection{Unambiguous Properties}
 Our main goal is to find  conditions under which 
 marginals give us enough information to identify desirable
allocations, irrespective of how agents extend their marginals to
preferences over bundles. In such cases, we say that the allocation
or mechanism satisfies a property \emph{unambiguously}. Formally, fix a
profile of marginal preferences $\succsim$. An allocation $\mu\in
\mathcal M$ is
\begin{itemize}
\item \textbf{unambiguously efficient} if for each preference profile $R$ that is responsive to $\succsim$, $\mu$ is efficient at $R$.
\item \textbf{unambiguously individually rational} if for each profile of preferences over bundles $R$ that is responsive to $\succsim$, $\mu$ is individually rational at $R$.
\end{itemize}

We now provide a strengthening of individual rationality. Given some profile of marginal preferences $\succsim$, an
allocation $\mu\in\mathcal M$ is \emph{component-wise individually
  rational} if, for any agent $i\in I$ and each of $i$'s endowed
objects $o\in\omega(i)$,  $|\{p\in\mu(i):p\succsim_i o\}|\geq |\{p\in\omega(i):p\succsim_i o\}|.$ It turns out that this seemingly stronger version of individual rationality is equivalent to unambiguous individual rationality.

\begin{proposition}\label{unambir}
  Given a profile of marginal preferences $\succsim$, an allocation
  $\mu$ is unambiguously individually rational if and only if it is
  component-wise individually rational.
\end{proposition}
\begin{proof}
    Let   $\succsim$ be a profile of marginal preferences and $\mu$ be an allocation. The ``if''-direction follows immediately since for responsive preferences, an agent always becomes weakly better off when we replace an object with another one that she marginally prefers. For the ``only if''-direction assume that $\mu$ is not component-wise individually rational. So, there are an agent $i$ and an object $o^*\in
  \omega(i)$ such that $K = |\{p\in \mu(i): p\succsim_i o^*\}| <
  |\{p\in \omega(i): p\succsim_i o^*\}| = L$. Let $T =
  |\omega(i)|$. We now construct a responsive extension $R_i$ of $\succsim_i$ for which $\mu(i)$ is not individually rational. Specifically, we produce an additive representation $u_i$ of such an $R_i$. To construct $u_i$, first assign some value $u_i(p)\in [0,1)$ to each
  $p\in O$ such that $p\succsim_i o^*$ (where $u_i(p)>u_i(q)$ if $p\succ_i q\succsim_i o^*$). Next, assign some value $u_i(r)\in (-T-1,-T)$ to each $r\in O$ such that $o^* \succ_i r$ (where $u_i(r)>u_i(s)$ if $o^*\succ_i r\succ_is$).
  Then $i$'s utility from $\omega(i)$ is at least  $-(T-L)(T+1)$.
  Similarly, $i$'s utility from $\mu(i)$ is at most $ -(T-K)T+K  =
  -T^2 +KT + K< -T^2 +(L-1)T + L = 
  -(T-L)(T+1)$. Thus, for $R_i$ represented by $u_i$,   $\omega(i)\mathrel
  P_i\mu(i)$, so $\mu$ is not unambiguously individually rational.  
\end{proof}
Before proceeding, note that we do not need to define an ``unambiguous'' version of strategy-proofness since marginal mechanisms are simply a subclass of mechanisms to which we can readily apply the usual definition of strategy-proofness. 

\section{Achieving Unambiguous Efficiency and Unambiguous Individual Rationality}\label{sec:compatibility IR E}

Marginal mechanisms force the designer to rely on nothing more than agents' rankings of individual objects to compute a full allocation. The next example shows that it is not always possible for such mechanisms to guarantee that an efficient and individually rational allocation is chosen.

\begin{example}\label{ex: simple} Let $I = \{1, 2\}$ and $\omega = (\{o_1, o_2\}, \{p_1, p_2\})$. Suppose both agents have the same marginal preferences, ranking $o_1$ first, $p_1$ second, $p_2$ third, and $o_2$ last. Giving $\{o_1, p_1\}$ to one agent and $\{o_2, p_2\}$ to the other is unambiguously efficient. We can be confident of this even if we only have information about marginal preferences. Moreover, knowing only these marginal preferences, we can also conclude that such an allocation is not individually rational: whoever gets $\{o_2,p_2\}$  prefers to keep their endowment no matter how their marginal preference extends to bundles. The designer needs to know more to achieve efficiency as well. Specifically, they need to know how   each agent compares the bundles $\{o_1, o_2\}$ and $\{p_1,  p_2\}$. Both ways of ranking these bundles are responsive to the marginal above since one swap makes her better off and the other makes her worse off. Thus, while it is straightforward to find allocations that unambiguously satisfy the requirements of efficiency and individual rationality separately, no single allocation satisfies both.
  \hfill $\circ$
\end{example}

There is a natural domain of marginal preferences for which unambiguously efficient and individually rational allocations do exist. A marginal preference $\succsim_i$ is \textbf{trichotomous} if it partitions the set of all objects $O$ into  \emph{attractive} objects $A_i$,  \emph{bearable} objects $B_i$, and  \emph{crummy} objects $C_i$ such that:
\begin{enumerate}
    \item $A_i, B_i,$ and $C_i$ are the indifference classes of $\succsim_i$,
    \item $\succsim_i$ ranks  $A_i$  above $B_i$ and  $B_i$ above  $C_i$, and
    \item No object in $i$'s endowment is crummy: $\omega(i) \cap C_i = \emptyset$.\footnote{In the context of shift exchange, $A_i$ would represent the endowed or non-endowed shifts for which $i$ experiences no scheduling conflict at all, $B_i$ would represent the endowed or non-endowed shifts for which $i$ experiences ``soft'' scheduling conflicts (e.g. an alternative activity that $i$ has not committed to or could cancel if necessary), and $C_i$ represents non-endowed shifts for which $i$ experiences ``hard'' scheduling conflicts (e.g. an alternative activity that $i$ has committed to and cannot cancel). Note that we assume that $i$ never experiences any hard scheduling conflicts with endowed shifts. One motivation for this assumption is that any hard scheduling conflicts have already been accommodated in the determination of the initial schedule (e.g. the unmissable wedding of a close relative is announced far enough in advance that the original schedule does not conflict with it).}
\end{enumerate}
This domain expands  the domain studied in \citet{ManjunathWestkamp:2018a}, to allow non-endowed objects in the set of bearable objects.
A trichotomous marginal $\succsim_i$ is fully identified by the attractive and bearable sets.\footnote{Recall that we assume endowments to be known by the designer.} So, we represent trichotomous marginal preferences by    $(A_i, B_i)$ 
or $\succsim_i$ interchangeably. We also denote the set of all trichotomous marginals for agent $i$ by $\mathcal{AB}_i$ and the set of all profiles of trichotomous marginals by $\mathcal{AB}$.

We now adapt the  Component-wise Individually Rational Priority (CIRP) algorithm  of \citet{ManjunathWestkamp:2018a} to compute an unambiguously efficient and unambiguously individually rational allocation. It starts with the set of all component-wise individually rational allocations and maximizes the welfare of agents, one at a time, in a fixed order. In order to formally define the algorithm, we fix a profile of desirable-bearable sets $(A,B)\in\mathcal {AB}$ and cinsider a fixed labelling of $I$ as $1, \dots, n$. Given this input, the CIRP algorithm proceeds as follows: 
 
\begin{description}
\item[\textbf{Step 0}:] Let $\mathcal{M}^0(A,B)$ be the set of all
  component-wise individually rational allocations at $(A,B)$. 

\item[\textbf{Step \boldmath $t\in\{1,\ldots,n\}$}:] 
Let 
\[
  K^t(A,B) = \max_{\mu\in \mathcal M^{t-1}(A,B)}
 |\mu(t)\cap A_t|
\]
be the \emph{promise to agent $t$} and 
\[
  \mathcal M^t(A,B) = \{\mu\in \mathcal M^{t-1} (A,B):
  |\mu(t)\cap A_t| = K^t(A,B)\}
\]
be the set of all allocations in $\mathcal M^{t-1}(A,B)$ that comply with
the promise to $t$.\footnote{\cite{ManjunathWestkamp:2018a} show how to compute the sequences $\{K^t(A,B),\mathcal M^t(A,B)\}_{t=1}^n$ in polynomial time.}
\end{description}

The arguments in \citet{ManjunathWestkamp:2018a} imply the following on the trichotomous domain.

\begin{proposition}\label{prop:exist E IR}
  For each  profile $(A,B)\in\mathcal{AB}$, every allocation in $\mathcal M^n(A,B)$ is
  unambiguously individually rational and unambiguously efficient.
\end{proposition}

While \cref{prop:exist E IR} establishes a sufficient condition for the compatibility of unambiguous efficiency and unambiguous individual rationality, we naturally ask whether this is true of any  bigger domains.  We start our answer to this question by disciplining what kinds of domains we  consider. Above, our definition of a trichotomous  marginal $\succsim_i = (A_i, B_i,C_i)$ restricts  which indifference classes can contain endowed objects ($A_i$ and $B_i$) and which indifference classes can contain non-endowed objects ($A_i$, $B_i$ and $C_i$). 
One way to encode such a marginal would be through a scoring function giving objects in  $A_i$ a score of 1, objects in $B_i$ a score of 0, and those in $C_i$ a score of $-1$. So, the restrictions to  trichotomous preferences tells us which scores an object can get based on whether it is an endowment or not.

We now consider a family of domains that can be described with similar language.
First, every marginal preference $\succsim_i\in \mathcal S_i$ can be represented by a scoring function $s_i:O\to \mathbb Z$ such that for each pair $o,p\in O$, $o\succsim_i p$ if and only if $s_i(o) \geq s_i(p)$.\footnote{Note that there can be multiple scoring functions that represent the same marginal. These scores only  encode an agent's ordinal rankings  are not meant to represent cardinal utilities.} Like our description of trichotomous preferences, we can constrain the scores that can be associated with endowments and non-endowments.
Let $\mathcal E$ be the  admissible scores for endowments and $\mathcal N$ be those for non-endowments. Then a marginal preference $\succsim_i\in \mathcal S_i$ is in the domain defined by $(\mathcal E, \mathcal N)$ if and only if it can be represented  by a scoring function $s_i$ such that for each endowment $o\in \omega(i), s_i(o)\in \mathcal E$ and for all other $o\in O\setminus \omega(i), s_i(o)\in \mathcal N$. We write $\mathcal S^{(\mathcal E, \mathcal N)}$ to represent this domain of marginals.

Effectively, the domain defined by $(\mathcal E, \mathcal N)$ only relies on restricting which indifference classes can contain endowed objects and which ones can contain non-endowed objects.
 Here are some concrete examples:
\begin{enumerate}
    \item Unrestricted: $\mathcal E = \mathcal N = \mathbb Z$
    \item Strongly trichotomous \citep{ManjunathWestkamp:2018a}: $\mathcal E =\{0,1\}$ and $\mathcal N = \{-1, 1\}$
    \item Trichotomous: $\mathcal E =\{0,1\}$ and $\mathcal N = \{-1,0,1\}$
\end{enumerate}
Note that the  $\mathcal E$-$\mathcal N$ description of a given domain of marginals is not unique. For example, we could describe the trichotomous domain by $\mathcal E' = \{1,10\} $ and $\mathcal N' = \{-10, 1, 10\}$. Nonetheless, our results pin down the \emph{ordinal} features of such domains as we describe below.

We impose 
a
requirement
 on the   $\mathcal E$-$\mathcal N$ pairs that we consider
that  
    non-endowments can take the extreme scores. That is, $\min \left(\mathcal E\cup \mathcal N\right)$, $\max \left(\mathcal E\cup \mathcal N\right)\in \mathcal N$.
This 
says that agents \emph{can} have an aversion to any non-endowed object (by giving it the lowest possible score) or a desire for it (by giving it the best possible score).   

We are now in a position to answer our question of when unambiguous efficiency and unambiguous individual rationality can simultaneously be achieved. Our next result gives a necessary condition on  $\mathcal E$-$\mathcal N$ domains for this to be the case.
\begin{theorem} \label{thm: max PE IR}Suppose there are at least three
  agents, each of whom is endowed with at least four objects.
  If $S^{(\mathcal E, \mathcal N)}$ admits  an unambiguously efficient and individually rational mechanism, then $\mathcal E$ contains no scores outside the two highest scores in $\mathcal E\cup \mathcal N$.
\end{theorem}

\cref{thm: max PE IR} gives us a necessary condition for there to exist unambiguously efficient and  individually rational allocations. As we see above, we can describe the trichotomous domain as an $\mathcal E$-$\mathcal N$ domain that  satisfies that condition and \cref{prop:exist E IR} tells us that such allocations exist on it. 
The only other domains of marginals that can be described by pairs of $\mathcal E$ and $\mathcal N$ satisfying the condition of \cref{thm: max PE IR} are ones that allow agents to have additional indifference classes below their top three containing only non-endowed objects. In other words, we \emph{could} start with a trichotomous marginal preference and break the crummy set into multiple indifference classes. Unambiguous individual rationality requires that no agent be allocated such an object, so whether these objects are split into more than one indifference class makes no material difference. Hence, we can interpret \cref{thm: max PE IR} and \cref{prop:exist E IR} as implying that the trichotomous domain is \emph{essentially} the unique maximal domain on which unambiguous efficiency and unambiguous individual rationality can be achieved simultaneously.

\begin{proof}[Proof of \cref{thm: max PE IR}]

Let $\sigma_1$ and $\sigma_2$ be the highest and second highest scores in $\mathcal E\cup \mathcal N$. To establish \cref{thm: max PE IR}, we prove a necessity statement: if the domain violates the condition that $\mathcal E$   
$\subseteq \{\sigma_1,\sigma_2\}$,  then the existence of an unambiguously efficient and individually rational allocation is not guaranteed. The argument proceeds by explicit construction. We consider all ways that the condition may be violated and, for each case, provide a preference profile for which no allocation is unambiguously efficient and individually rational. 

Suppose to the contrary that unambiguous individual rationality and unambiguous efficiency are compatible despite there being some $k>2$ such that $k^{\text{th}}$ highest score $\sigma_k\in \mathcal E$. We first consider the possibility that $\sigma_2\in \mathcal N$, meaning that agents can rank non-endowments in their second indifference  class. We consider  the remaining possible case where $\sigma_2\notin \mathcal N$ afterwards. 

It is without loss of generality to focus on the case where there are exactly three agents and each agent is endowed with exactly four objects.\footnote{Recall that agents can always assign the lowest possible score to any non-endowed object, thus preventing any unambiguously individually rational allocation from assigning them that object.}  Let $\omega(i) =\{o_i,p_i,p'_{i},q_i\}$. We will refer to $\{o_1,o_2,o_3\}$ as ``$o$-objects'', to $\{p_1,p'_1,p_2,p'_2,p_3,p'_3\}$ as ``$p$-objects'', and to $\{q_1,q_2,q_3\}$ as ``$q$-objects''. Since $\{\sigma_1,\sigma_2\}\subseteq \mathcal{N}$ and $s_k\in\mathcal{E}$, the following profile of marginal preferences is in the domain we currently consider: 
\[
  \begin{array}{cccl}
    \text{Score}&\succsim_i\\
    \hline
    \sigma_1&o_j & \forall j \in I\setminus\{ i\} & \text{(Others' $o$-objects)} \\ 
    \sigma_2  &p_j,p'_{j} & \forall j \in I\setminus\{i\}&\text{(Others' $p$-objects)}\\
    \sigma_k &o_i, p_i, p'_i, q_i& & \text{Endowment}
  \end{array}
\]

If $\mu$ is unambiguously individually rational, then for each pair
$i, j\in I$ such that $i\neq j, \mu(i)\not\ni q_j$. So, by
feasibility, $q_i\in \mu(i)$. That is, each agent keeps her
$q$-object.

Next, we show that each agent $i\in I$ gets exactly one $o$-object and that this $o$-object was not in $i$'s endowment. Assume first that, without loss of generality given symmetry,  $1$  keeps $o_1$. If $\mu(j)$ contains $o_j$ for $j\in\{2,3\}$, then $1$ and $j$ become strictly better off by trading their $o$-objects. Hence, we must have $o_2\in\mu(3)$ and $o_3\in\mu(2)$. By individual rationality and feasibility, $\mu(2)$ and $\mu(3)$ must both contain two $p$-objects. Furthermore, either $\mu(2)$ or $\mu(3)$ must contain a $p$-object that did not initially belong to $i$. Exchanging that $p$ object for $o_1$ will make both agents strictly better off. Hence, it cannot be that $1$ or any other agent keeps her $o$-object. Next, suppose - again without loss of generality given symmetry - that $1$ gets $o_2$ and $o_3$, while $2$ receives no $o$-objects at all. Since $2$ receives no
$o$-objects and can't trade away her $q$-object, by feasibility and unambiguous individual rationality, $2$ must receive exactly three $p$-objects. If two of these $p$-objects did not initially belong to $1$, then consider the trade between $1$ and
$2$ where $1$ gives $o_3$ and $q_1$ to
$2$ in return for any two $p$-objects in $\mu(2)\setminus \omega(i)$. Clearly, there is a responsive extension $R$ of the marginal preference profile $\succsim$ such that the just described trade makes $1$ and $2$ strictly better off. If $2$ has both of $i$'s $p$-objects, then she could first swap one of these objects for another $p$-object in $\mu(3)$ (this would clearly make $3$ at least weakly better off) and then trade as in the previous case with $1$.  Hence, no agent receives more
than one  $o$-object in an unambiguously efficient and unambiguously individually rational allocation. Together with our assertion that no agent keeps her $o$-object, we obtain that every agent receives exactly one
$o$-object that was initially endowed to some other agent.

Since each agent keeps their own $q$-object and receives exactly one
other agent's $o$-object, feasibility of $\mu$  and unambiguous individual rationality dictate that every
agent receives exactly two $p$-objects. If $i$ keeps $p\in \{p_i,
p'_i\}$, then for 
any $j\in I\setminus \{i\}$, if $i$ trades $p$ to $j$ for a
$p$-object in $\mu(j)\setminus \omega(i)$, then $i$ is better off while $j$ is
unaffected or better off, contradicting efficiency. 

Summing up our arguments above, we see that unambiguous efficiency and unambiguous individual rationality imply that each agent $i$ gets:
\begin{enumerate}
\item The $o$-object of some $j\neq i$.
\item Two $p$-objects that initially belonged to $j,k\neq i$ (where we allow for the case where $j=k$).
\end{enumerate}
We now derive a contradiction to unambiguous efficiency. Assume to the contrary that there is an unambiguously efficient and unambiguously  individually rational allocation $\mu$. Consider Agent $1$ and assume, without loss of generality, that $o_2\in \mu(1)$. If both of Agent $3$'s $p$-objects in $\mu(3)$ did not initially belong to $1$, then a trade in which $1$ and $3$ swap $\{o_2,q_1\}$ and the two $p$-objects in $\mu(3)$ makes both agents better off for some responsive extension $R$ of the marginal preference profile $\succsim$. Otherwise, Agent $3$ could first swap all $p$-objects in $\mu(3)$ that initially belonged to $1$ with $2$ for $p$-objects in $\mu(2)$ that did not initially belong to $1$ (this would not affect the welfare of $2$) and then trade as in the previous case with $1$. 

Next, we consider the alternative possibility that \aw{$\sigma_2\notin \mathcal N$} meaning that agents can only rank endowments in their second indifference  class. As above, it is without loss of generality to suppose that $I=\{1,2,3\}$ and that two agents are endowed with four objects each and
the third is endowed with three.
 Let $\omega(1) = \{o_1, o_2, o_3,
q_1\}$,$\omega(2) =   \{p_1, p_2, p_3, q_2\}$, and $\omega(3) =
\{a_1,a_2, a_3\}$. 
  By assumption that $\sigma_k \in \mathcal E$, endowments are \aw{allowed in some indifference class
  below the second one} but since $\sigma_2 \notin \mathcal N$, non-endowments cannot be ranked in the
  second indifference class. So, we can find a profile of marginal
  preferences that is ordinally equivalent to the following:
  \[
    \begin{array}{ccc}
      \succsim_1 &\succsim_2&\succsim_3\\
      \hline
      a\text{-objects} & a\text{-objects} &
                                            o\text{-objects},p\text{-objects}\\
      o\text{-objects} & p\text{-objects} & a\text{-objects}\\
      q_1 & q_2 & 
    \end{array}
  \]
At any unambiguously  individually rational allocation, 1 and 2 each
keep their own $q$-object while 3 does not get a $q$-object. Moreover,
1 does not get any $p$-objects and 2 does not get any $o$-objects.
At any efficient allocation, 3 does not keep any $a$-objects, so they
are split between 1 and 2. Thus,
if $\mu$ were an unambiguously efficient and unambiguously individually rational
allocation, then for some $k=1,\dots, 3$, agent 1 gets $k$
$a$-objects, $3-k$ $o$-objects, and one $q$-object, agent 2 gets $3-k$
$a$-objects, $3-(3-k)$ $p$-objects, and one $q$-object, and agent 3 gets
$k$ $o$-objects and $3-k$ $p$-objects. Without loss
of generality, suppose $k \geq 2$. Let (again without loss of generality) $a_1\in \mu(1), p_1,p_2
\in \mu(2)$, and $o_1, o_2 \in \mu(3)$. Consider $\nu$ achieved by
starting from $\mu$ and having 1 give 2 $\{a_1,q_1\}$, 2 give 3
$\{p_1, p_2\}$ and 3 give 1  $\{o_1, o_2\}$.
Since $a_1 \succ_1 o_1$ and $o_2 \succ_1 q_1$, there is a responsive extension of $\succsim_1$ such that $1$ strictly prefers $\nu(1)$
over $\mu(1)$. Since $a_1 \succ_2 p_1, p_2 \succ_2 q_1$, there is a responsive extension of $\succsim_2$ such that $2$ strictly prefers $\nu(2)$
over $\mu(2)$.  Finally, agent 3 is indifferent between $\nu(3)$ and
$\mu(3)$. Hence, there is a responsive extension for which $\nu$ welfare-dominates $\mu$ and $\mu$ is not unambiguously efficient.  
\end{proof}

\section{Incentives}\label{sec:SP}
\label{sec: incentives}
On the trichotomous domain, the CIRP algorithm always finds an unambiguously efficient and unambiguously individually rational allocation. What happens when we add strategy-proofness as a third desideratum? We first show that the strongly trichotomous domain is maximal for the existence of mechanism that satisfies all three desiderata and then turn to a deeper analysis of incentives on the trichotomous domain.

\subsection{Maximality of the Strongly Trichotomous Domain}
\label{sec: sp max}

On the \emph{strongly} trichotomous subdomain, \citet{ManjunathWestkamp:2018a}  show that CIRP mechanisms are not only unambiguously efficient and unambiguously individually rational, but also strategy-proof.

\begin{theorem}[\cite{ManjunathWestkamp:2018a}]
Every CIRP mechanism on the strongly trichotomous domain is strategy-proof.
\end{theorem}

We show next that this positive result does not extend to the trichotomous domain. Rather, the strongly trichotomous domain is the essentially unique maximal domain for which an unambiguously efficient, unambiguously individually rational, and strategy-proof mechanism exists.\footnote{See the discussion preceding the proof of \cref{thm: max PE IR} for what we mean by ``essentially unique''.} In fact, our proof  implies that no efficient, individually rational, and strategy-proof mechanism, marginal or not, exists beyond the strongly trichotomous domain.

\begin{theorem}\label{thm:SPnec}
Suppose there are at least four agents, at least one of whom is
endowed with two or more objects. If $S^{(\mathcal E, \mathcal N)}$ admits an unambiguously efficient, unambiguously individually
rational, and strategy-proof mechanism, then $\mathcal E$ contains no scores outside the two highest scores and $ \mathcal N$ does not contain the second highest score.
\end{theorem}

\begin{proof}[Proof of \cref{thm:SPnec}]

  Given Theorem~\ref{thm: max PE IR}, it suffices to prove that if non-endowments are allowed in the second indifference class, the three properties are incompatible. In the proof, we consider a situation where only one agent is endowed with two objects and the rest are endowed with just one.  With just this minimal variation from the \citet{ShapleyScarf:JME1974} model, we arrive at an impossibility even though all agents have trichotomous preferences.

Let four of the agents be labelled 1, 2, 3, and 4. We start with the
case where agent 1 is endowed with object $o$, agent 2 is endowed
with $p$, agent 3 is endowed with  $\{q_1, q_2\}$, and agent 4 is
endowed with $r$. We can  extend the argument below to larger
endowments in the same way as the argument in the proof of
Theorem~\ref{thm: max PE IR} by letting the remaining endowments  be  \A for
their owner but unacceptable to all other agents. 

The following marginal preference  profile is in the domain:
\[
\begin{array}{cccc}
  \succsim_1 & \succsim_2 & \succsim_3 & \succsim_4\\
  \hline
  q_1 & q_1& o,p & q_2 \\
  o, r & p, r & q_1, q_2  & r \\ 
\end{array}
\]
We argue first that in any unambiguously efficient and unambiguously individually rational allocation, 3 must get $o$ and $p$. If $3$ gets
neither, $3$ must retain both of her assigned objects. Unambiguous individual rationality then requires that the three other agents must all consume their own endowments as well.
However, $3$ would benefit by trading $q_1$ with either $1$ or $2$ for their endowment. If $3$ gets only one of
$o$ or $p$, then by unambiguous individual rationality, $3$ keeps one of her endowed objects. Without loss of generality, suppose that $3$ gets $o$. If the
other object $3$ gets is $q_1$, then 1 must be consuming $r$, 4
must be consuming $q_2$ and 2 must be consuming $p$. Then 3 could
trade $q_1$ for $p$ with 2, thereby making both agents strictly better off. Suppose instead that 3 gets $o$
and $q_2$. Then, since no other agent finds $p$ acceptable, 2 must be
consuming it. Since 1 is the only other agent who finds $q_1$
acceptable and 3 is not consuming it, 1 must be consuming
$q_1$. Therefore 4 must be getting $r$. The three-way trade where 3
gives $q_2$ to 4, 4 gives $r$ to 2, and 2 gives $p$ to 3 is
welfare-improving. Thus, 3 must be getting both $o$ and $p$. Individual
rationality then requires that 4 gets $q_2$ since she is the only agent
who finds $q_2$ acceptable. So, each of 1 and 2 gets either $q_1$ or 
$r$. Thus, there are two unambiguously efficient and unambiguously individually rational
allocations:

\[
\begin{array}{ccccc}
  & 1 & 2 & 3 & 4\\
  \hline
  \mu^1& q_1 & r & o,p & q_2\\
  \mu^2& r & q_1 & o,p & q_2\\
\end{array}
\]

The problem is symmetric and without loss of generality suppose a rule
picks $\mu^1$.
Now consider a new profile of preferences where only 2's preferences
are changed.

\[
\begin{array}{cccc}
  \hline
  \succsim_1 & \succsim_2' & \succsim_3 & \succsim_4\\
  \hline
  q_1 & q_1& o,p & q_2 \\
  o, r & p & q_1, q_2  & r \\ 
\end{array}
\]

By strategy-proofness, 2 cannot get $q_1$. So, by individual
rationality, they get $p$. If 3 does not get $o$, then they keep their
two endowments and 1 keeps $o$. Then, 3 and 1 can benefit by trading
$q_1$ for $o$, violating efficiency. Thus, 3 gets $o$ and keeps
one of $q_1$ or $q_2$. If 3
keeps $q_1$, then individual rationality says that 4 gets $q_2$ and 1
gets $r$. If 3 keeps $q_2$, then individual rationality says that 4 gets $r$ and 1 gets
$q_1$. Thus, there are two possible allocations a rule could choose.
\[
\begin{array}{ccccc}
  \hline
  & 1 & 2 & 3 & 4\\
  \hline
  \mu^3& q_1 & p & o, q_2 & r\\
  \mu^4& r & p & o,q_1 & q_2\\
\end{array}
\]
In either case, 3 gets only one object, $o$, in their top indifference
class. So we consider another problem where only 3's preferences
change. 
\[
\begin{array}{cccc}
  \hline
  \succsim_1 & \succsim_2' & \succsim_3' & \succsim_4\\
  \hline
  q_1 & q_1& p & q_2 \\
  o, r & p & q_1, q_2,o  & r \\ 
\end{array}
\]
If 3 gets $p$, then individual rationality says that 2 gets $q_1$. If
3 additionally gets $q_2$, then individual rationality says that 4
gets $r$ and 1 gets $o$. However, the three-way trade between 1, 3,
and 4 where 1 gets $r$, 3 gets $o$, and 4 gets $q_2$ is
welfare-improving. So, if 3 gets $p$ then they must also get $o$,
violating strategy-proofness regardless of whether the rule chooses
$\mu^3$ or $\mu^4$ in the above problem. Thus, 3 does not get $p$. If
they do 
not get $o$, then they keep their endowment and again individual
rationality says that all of the other agents keep their endowments as
well. But the same three-way trade between 1, 3, and 4 is
welfare-improving, contradicting efficiency. Thus, 3 gets $o$
and keeps one of their endowments. Individual rationality says that if
3 keeps $q_1$ then 4 gets $q_2$, 1 gets $r$, and 2 gets $p$. But then, $2$ and $3$ could trade $q_1$ and $p$ making both of them (unambiguously) strictly better off. Hence, $3$ must keep  $q_2$ and the only efficient individually rational allocation is $\mu^3$ from above.  

Now consider one more preference profile where we change 3's
preferences again.

\[
\begin{array}{cccc}
  \hline
  \succsim_1 & \succsim_2' & \succsim_3'' & \succsim_4\\
  \hline
  q_1 & q_1& p & q_2 \\
  o, r & p & q_1, q_2 & r \\ 
\end{array}
\]
Strategy-proofness says that $3$ cannot receive $p$ and must hence keep $q_1$ and $q_2$. Individual rationality says that all agents then keep their endowments. However, this contradicts efficiency since 3 could trade $q_1$ to 2 in exchange for $p$ and they would both be strictly better off.
\end{proof}

Note that in the proof above, we do not rely on different responsive extensions of the marginal preferences we consider. Hence, our proof  implies that no mechanism, marginal or not, is efficient, individually rational, and strategy-proof on the trichotomous domain. 

\subsection{Incentives on the Trichotomous Domain}\label{sec:incentives trichotomous}
As we have seen above, the CIRP mechanism satisfies our allocative criteria on the trichotomous domain but is not strategy-proof on that domain. Intuitively, the reason is that agents can never benefit and are often disadvantaged by revealing non-endowed objects in their \B set. This is reminiscent of the issues that arise with compatible patient-donor pairs in the kidney exchange problem (see, e.g., \citet{sonmezyenmez20} and the references therein). Interestingly, CIRP mechanisms still have a weakly dominant strategy on the trichotomous domain that involves ``hiding'' any non-endowed objects in an agent's middle indifference class. 

\begin{theorem}\label{prop: weak dom truth}
    On the trichotomous domain,    under any CIRP mechanism, it is a weakly dominant strategy for each agent to report their \A set truthfuly and report only endowed \B objects. That is, for each profile $(A,B)\in \mathcal {AB}$, each $i\in I$, and each $(A_i', B_i') \in \mathcal{AB}_i$, for each $\mu\in \mathcal M^n((A_i, \omega(i)\setminus A_i), (A_{-i},B_{-i}))$ and $\mu'\in \mathcal M^n((A_i', B_i'), (A_{-i},B_{-i}))$ we have $\mu_i \mathrel R_i \mu_i'$ for any responsive $R_i$ with marginal trichotomous preference $(A_i,B_i)$.
\end{theorem}
While intuitive, the proof of this result is surprisingly involved. We provide a brief sketch below and the complete proof is available in \cref{proof of weak dom truth}.

\paragraph{Proof sketch} 
Fix a profile $(A,B)\in\mathcal{AB}$ and an agent $i$. The argument reduces an arbitrary report $(A_i',B_i')$ to the benchmark report $(A_i,\omega(i)\setminus A_i)$ via a sequence of single-object edits, where in each edit one object is moved between the three trichotomous categories \A, \B, and \C. For each such edit, we compare the sets of allocations produced by the CIRP procedure before and after the edit.

The key technical step is a tight \emph{one-object, one-cycle} sensitivity property: when a single object is moved between $i$'s categories, the induced change in the CIRP outcome can be represented in a symmetric-difference graph that decomposes into alternating cycles, and at most one cycle can be affected. Consequently, a single-object report change can alter $i$'s final assignment only through a single cycle and therefore can change $|\mu(i)\cap A_i|$ by at most one, with a correspondingly tight description of which object(s) move in or out of $i$'s bundle.

With this one-cycle sensitivity in hand, the dominance conclusion is obtained by a  bookkeeping argument. First, deleting a non-endowed bearable object from the report, that is, moving it from $B_i'$ to $C_i'$, cannot make $i$ worse off and, iterating, $i$ does not lose from restricting her reported \B set to endowed \B objects (that is, objects in $\omega(i)\setminus A_i$). Second, misreporting in the \A dimension has tightly limited upside: moving a single object into or out of the reported \A set can change the number of true \A objects obtained by at most one, and any such change is offset by the possibility of giving up a true \A object along the unique affected cycle. Iterating these one-object comparisons yields that $(A_i,\omega(i)\setminus A_i)$ weakly dominates any alternative report $(A_i',B_i')$. 

\bigskip

\cref{prop: weak dom truth} allows us to characterize the unique Nash equilibrium of CIRP in weakly dominant strategies. 
Moreover, our subsequent result (\cref{cor:weak eff}) shows that such an equilibrium allocation is guaranteed to be \emph{weakly} efficient with respect to true preferences. However, since trichotomous marginals are so coarse,  \emph{weak}  efficiency is much weaker than efficiency.  In this sense, CIRP is an unsuitable solution in such settings since truth-telling is weakly dominated and equilibrium is inefficient.

Given the importance of  efficiency in applications, we ask whether we can do better in terms of  balancing efficiency, individual rationality, and incentives on the trichotomous domain.
We provide a novel algorithm that provides agents with better incentives to reveal their middle indifference class truthfully and ensures that truth-telling is not weakly dominated. 

Our algorithm uses a series of serial dictatorships on the set of
component-wise individually rational allocations that only relies on
which non-endowed objects are in an agent's \B set once it can no
longer improve that agent (by giving her more of her \A objects). To formally define our algorithm, we first augment the definition of the CIRP algorithm with a "current" allocation $\mu$:
\begin{description}
\item[ \textbf{Step 0:}] Let $\mathcal{M}^0(A,B, \mu)$ be the set of all
  component-wise individually rational allocations at $(A,B)$ that
  weakly welfare-improve on $\mu$ for all agents.

\item[\textbf{Step \boldmath $t\in\{1,\ldots,n\}$:}] 
Let 
\[
  K^t(A,B,\mu) \equiv \max_{\mu\in \mathcal M^{t-1}(A,B,\mu)}
 |\mu(t)\cap A_t|
\]
be the \emph{promise to agent $t$} and 
\[
  \mathcal M^t(A,B,\mu) = \{\mu\in \mathcal M^{t-1} (A,B,\mu):
  |\mu(t)\cap A_t| = K^t(A,B,\mu)\}
\]
be the set of all allocations in $\mathcal M^{t-1}(A,B,\mu)$ that comply with
the promise to $t$.
\end{description}\medskip

In words, given a  profile of \A and \B sets $(A,B)$ and a  component-wise individually rational matching $\mu$, $\mathcal M^t$ is the set of allocations that remain after the $t^{\text{th}}$ round of a serial dictatorship (where agents take turns in increasing order of their indices) over the set of component-wise individually rational allocations that welfare-dominate $\mu$. 

Our  algorithm uses these functions  on a sequence of profiles of \A
and \B sets. Initially, we run the serial dictatorship algorithm
assuming that agents may only be matched to objects in their \A sets
and their endowments. As we show below, there is at least one agent,
who can no longer improve (receive strictly more \A objects)
at this  serial dictatorship outcome, \emph{irrespective} of which
objects other agents find \B. Each such agent is then asked to reveal
her set of \B objects and we compute the resulting serial dictatorship outcome. We iterate this procedure until no improvable agents remain and we have elicited all \B sets. 

Formally, set $\mu^0 = \omega$ and, for each $i\in I$, $B_i^0 = \omega(i)\setminus A_i$ and $\overline{B}_i^0=O\setminus A_i$. Now, for each $t\in \{1, \dots, n\}$, let  $\mu^t\in \mathcal
M^n(A,B^{t-1},\mu^{t-1})$, let
\[
  I^t = \{i\in I: \nexists\mu\in \mathcal M^0(A,\overline B^{t-1},\mu^t)\text{
    s.t. }|\mu(i)\cap A_i| > |\mu^t(i)\cap A_i| \},
\]  
let  $B^t$ be such that for each
$i\in I$,
\[
  B_i^t = \left\{
    \begin{array}{ll}
      B_i&\text{ if }i\in I^t\\
      B_i^0 & \text{  otherwise},
    \end{array}
    \right.
  \]
  and let  $\overline B^{t}$ be such that for each $i\in I$,
  \[
    \overline B_i^{t} = \left\{
      \begin{array}{ll}
        B_i&\text{ if }i\in I^t\\
        \overline B_i^0 & \text{  otherwise}.
    \end{array}
    \right.
  \]

We next establish that this \emph{gradual} revelation of unimprovable agents' \B sets is possible, and that the result is unambiguously efficient and unambiguously individually rational.
\begin{theorem}\label{thm:sufficiency IR PE}
  For each  profile $(A,B)\in\mathcal{AB}$, let $(\mu^t,B^t,\bar B^t, I^t)$ be the four-tuple of sequences produced by the algorithm above.
\begin{enumerate}
\item There exists a $T$ such that $I^T = I$ and, for each $t< T, I^t \subsetneq
I$.
\item Any $\mu \in\mathcal M^n(A,B,\mu^T)$ is unambiguously efficient and unambiguously individually rational. 
\end{enumerate}
\end{theorem}

Before we present the somewhat involved proof of \cref{thm:sufficiency IR PE}, we now discuss the sense in which gradual revelation provides better incentives for truth-telling on the trichotomous domain than the straightforward extension of the canonical strategy-proof mechanism for the strongly trichotomous domain. We denote by $\varphi$ any mechanism that complies with the principles of gradual revelation mentioned above (any mechanism which, for any $(A,B)$, selects a matching from $\mathcal M^n(A,B,\mu^T)$). We have the following. 

\begin{theorem}\label{thm:grincentives}
Any gradual revelation mechanism $\varphi$ is
\begin{enumerate}
\item \emph{truncation proof}, that is, for any $(A,B)\in\mathcal{AB}$, any $i\in I$,  any $R_i$ 
responsive to $(A_i,B_i)$,
and any
$\hat{B}_i$ such that $\omega(i)\setminus A_i\subseteq \hat{B}_i\subseteq O\setminus A_i$, $$\varphi(A,B)\mathrel R_i\varphi(A,(\hat{B}_i,B_{-i})).$$
\item \emph{not obviously manipulable}, that is, for any $(A,B)\in\mathcal{AB}$, any $i\in I$, any $R_i$ responsive to $(A_i, B_i)$,  and any $(\hat A_i,\hat B_i)\in \mathcal{AB}_i$,\footnote{The
  max and min operators pick the best and worst outcomes for $i$ with
  respect to $R_i$,
  respectively.} 
     $$\max_{(A_{-i},B_{-i})}\varphi ((A_i,B_i),(A_{-i},B_{-i}))\mathrel
R_i\max_{(A_{-i},B_{-i})}\varphi ((\hat
A_i,\hat B_i),(A_{-i},B_{-i}))$$
and $$\min_{(A_{-i},B_{-i})}\varphi ((A_i,B_i),(A_{-i},B_{-i}))\mathrel
R_i\min_{(A_{-i},B_{-i})}\varphi((\hat
A_i,\hat B_i),(A_{-i},B_{-i})).$$
\end{enumerate}
\end{theorem}

\begin{proof}
\begin{enumerate}
    \item By construction, gradual revelation mechanisms only use information about non-endowed objects in an agent's middle indifference class, once that agent's number of \A objects can no longer be increased. Furthermore, from one iteration to the next, gradual revelation mechanisms only implement reallocations that make all agents weakly better off. Hence, an agent's revelation of her $B$ set can only influence which specific objects she gets from her top two indifference classes but not the number of objects she gets from either class. 
    \item Since gradual revelation mechanisms are truncation proof, any potentially profitable
manipulation $(\hat A_i,\hat B_i)$ must have $\hat A_i\neq A_i$. By
unambiguous individual rationality, the worst outcome for $i$ is
always to be stuck with all of their endowed objects. The best
possible outcome is for $i$ to get $\min\{|A_i|,|\omega(i)|\}$ of her
\A objects and clearly that best possible scenario is better for $i$
when their reported $A$ set is truthful.\footnote{Note that these arguments are also easily seen to imply that there is no untruthful preference report that weakly dominates truth-telling.}
\end{enumerate}
\end{proof}

Truncation-proofness has played a prominent role in the literature on two-sided matching markets, (see, for instance, \citet{rothrothblum1999} and \citet{ehlers2008}). By Theorem~\ref{thm:SPnec}, truncation proofness implies that gradual revelation mechanisms cannot make it part of a weakly dominant strategy for agents to reveal their \A sets truthfully. However, in contrast to CIRP mechanisms, potentially profitable manipulations in which an agent shrinks or expands their \A set cannot be weakly dominant since we can always find a profile of other agents' preferences for which the manipulation does strictly worse than truthful revelation. Furthermore, while profitable deviations from truth-telling can exist, these can never be obvious in the sense of \cite{troyanmorrill}.\footnote{There are important real-world mechanisms that are obviously manipulable. For example, \citet{troyanmorrill} show that the well-known Boston, or Immediate Acceptance, Mechanism is obviously manipulable.} The combination of the two limited incentive compatibility notions above suggests that gradual revelation mechanisms could be a useful new tool for solving balanced exchange problems on the trichotomous domain, the maximal domain for which marginal preferences are sufficient to unambiguously identify efficient and individually rational allocations.

\paragraph{Proof roadmap for \Cref{thm:sufficiency IR PE}.}
The termination claim in part (i) is driven by a monotonicity argument: once an agent becomes unimprovable in terms of additional \A objects, subsequent revelations of other agents' \B sets cannot make her improvable again. The efficiency and individual rationality claim in part (ii) follows from  a cycle-based characterization of welfare improvements among component-wise individually rational allocations. The remainder of this section  develops this cycle language and the supporting lemmas.

We need some auxiliary terminology and results to prove Theorem \ref{thm:sufficiency IR PE}. Given a matching
$\mu\in\mathcal M$, a \emph{cycle} of $\mu$ consists of a sequence of distinct agents,
$(i_l)_{l=1}^L$, and a sequence of distinct objects, $(o_l)_{l=1}^L$,
such that for each $l\in \{2,\dots, L\}$, $o_{l-1} \in \mu(i_l)$ and
$o_l\notin \mu(i_l)$, and $o_L\in \mu(i_1)$. Let $I(C)=\{i^1,\ldots,i^L\}$ be the set of agents involved in
$C$ and let $O(C)= \{o^1,\ldots,o^L\}$ be the set of objects involved in
$C$. We denote by $\mu+C$ the matching obtained by replacing, for each
$l\in\{2,\dots, L\}, o_{l-1}$ in $\mu(i_l)$ with $o_l$, and $o_{L}$ in
$\mu(i_1)$ with $o_1$. An important observation is that balancedness implies that we can mechanically move between any
pair of allocations by reallocating no object more than once. Given a pair $\mu, \nu\in \mathcal M$, we say
that a collection of cycles of $\{C_1, \dots, C_K\}$ is a
\emph{decomposition of $\nu-\mu$} if
  \begin{enumerate}
  \item $C_1,\dots,C_K$ are disjoint in terms of
  objects---that is, for each pair $l,k\in \{1, \dots, K\}, O(C_l)\cap O(C_k) =
  \emptyset$, and
\item executing all of the cycles, starting at $\mu$, yields
  $\nu$---that is,   \[
    (((\mu + C_1) + C_2)+\dots)+C_K = \nu.\footnotemark
  \]\footnotetext{Since the cycles are disjoint in terms of objects,
    the order of execution does not matter.}
  \end{enumerate} The next lemma from \citet{ManjunathWestkamp:2018a} states that such a decomposition exists for any pair of
allocations.
\begin{lemma}
  \label{lemma: decomposition}
  For each pair $\mu, \nu\in \mathcal M$, there is a decomposition of $\nu-\mu$.
\end{lemma}

Next, we define some additional properties of cycles based on agents'
preferences. We say that $C$ 
\begin{itemize}
\item is \emph{component-wise individually rational} if, for each $m\in\{
  1,\dots, M\}$, $o^m\in B_{i^m}\cup  A_{i^m}$,  
\item \emph{increases $i$'s welfare} if
  $|(\mu+C)(i)\cap A_i|\mathrel{>}|\mu(i)\cap A_i|$, 
  
  \item \emph{decreases $i$'s welfare} if
  $|(\mu+C)(i)\cap A_i|\mathrel{<}|\mu(i)\cap A_i|$,
  
  \item \emph{affects $i$} if it either increases or decreases $i$'s
    welfare, and
  \item is \emph{welfare-improving} if  it increases some agent's welfare
    without decreasing that of any other agent.

\end{itemize}

The next lemma establishes that we can characterize the efficiency of component-wise individually rational matchings via the absence of certain cycles. 

\begin{lemma}\label{prop:CIRP Pareto}
If $\mu$ is component-wise individually rational at $(A,B)$, then $\mu$ is
unambiguously efficient if and only if there is no component-wise
individually rational and welfare-improving cycle of~$\mu$.
\end{lemma}
\begin{proof}
  The ``only if'' part is trivial so we only establish the ``if'' part.
  Fix a component-wise individually rational allocation $\mu$ at   $(A,B)$. Assume that there is a matching $\nu$ that welfare-dominates   $\mu$. 
  By \cref{lemma: decomposition}, there is a decomposition $\{C_1, \dots, C_K\}$ of
  $\nu-\mu$. We say that $C'$ is a subcycle of $\nu-\mu$, if there exists a
  decomposition $\{C_1,\dots,C_K\}$ of $\nu-\mu$ such that $C_k=C'$ for some
  $k$.\footnote{There may be more than one decomposition
    of $\nu-\mu$.} If $C'=(i^1,o^1,\dots,i^M,o^M)$ is a subcycle of $\nu-\mu$, we let  
  \[
  \nu-C'=\nu+(i^1,o^M,i^M,o^{M-1},\dots,i^2,o^1).
  \] Note that $\nu-C'$ reverses the trades in $C'$ so that the
  associated objects return into the possession of those agents who
  owned them at $\mu$. Finally, we say that $\nu$ is a minimal
  welfare-improvement of $\mu$ if there is no non-empty subcycle $C'$ of $\nu-\mu$ such
  that $\nu-C'$ welfare-improves upon $\mu$.  
  
  To complete the proof, we establish that it is without loss of
  generality to assume that there is a cycle $C$ such that $\mu+C = \nu$. First,  it is
   without loss to assume that $\nu$ is a minimal
  welfare-improvement. Next, we  use $\mu$ and $\nu$ to construct a directed graph. Let $$I' =     \{i: i\in I\text{ and } \mu(i) \neq \nu(i)\} $$ and 
$$O' =   \{o: \text{there are } i,j\in I'\text{ such that }i\neq j \text{ and }o\in
  \mu(i)\cap\nu(j)\},$$
and let the set of vertices be $I'\cup O'$. For each $o\in O'$ and $i\in I'$ such that $o\in
\mu(i)\setminus \nu(i)$, insert the edge $(o,i)$. For each $i\in I'$, let
$o^i\in O'$ be one of $i$'s most preferred objects in $\nu(i)\setminus
\mu(i)$, meaning  $o^i\in \nu(i)\setminus \mu(i)$ and, for each
$o'\in \nu(i)\setminus \mu(i)$, $o^i\mathrel \succsim^*_i o'$,
and insert the edge $(i, o^i)$. Let $G$ denote the resulting directed graph. 
  
   Each vertex in $G$ has a single
out edge and there is no edge from any $i\in I'$ to $o\in
\mu(i)$. Therefore, $G$ has some cycle $C = (j^1, p^1, \dots, j^K, p^K)$ that involves at least two agents. 

We now show that $C$ weakly increases the welfare of all agents who are involved in it. For any $k$ such that $p^k\in A_{j^k}$, $C$ weakly increases the welfare of $j^k$. Now consider some $k$ such that $p^k\in O\setminus A_{j^k}$. Since $p^k$ is one of $j^k$'s most preferred objects in $\nu(j^k)\setminus \mu(j^k)$, we must have $\nu(j^k)\setminus \mu(j^k)\subseteq
O\setminus A_{j^k}$. Since $\nu(j^k)\succsim\mu(j^k)$ and since agent $j^k$'s preferences are trichotomous, we must have $\mu(j^k)\setminus \nu(j^k)\subseteq O\setminus A_{j^k}$ as well. Furthermore, given that $\mu$ is component-wise individually rational, we obtain $\mu(j^k)\setminus \nu(j^k)\subseteq B_{j^k}$ and $\nu(j^k)\setminus \mu(j^k)\subseteq
B_{j^k}$. Hence, $C$ must leave $j^k$ unaffected.

If $C$ is not a welfare-improving cycle of $\mu$, then it does not
affect any agent. However, this means
$\nu-C$ is a 
welfare-improvement of $\mu$,  contradicting the minimality of
$\nu$. Hence, $C$ is Pareto-improving and by the minimality of $\nu$,  $\nu=\mu+C$. 
\end{proof}

The final auxiliary result for the proof of Theorem~\ref{thm:sufficiency IR PE} is that while an expansion of agents' \B sets permits
Pareto-improvements, such an expansion can never yield a \emph{strict}
welfare-improvement.\footnote{This is reminiscent of an analogous result \citep{Erginetal2020} in the context of pairwise liver exchange.}
  
\begin{lemma}\label{lemma:noimprovement} Let $(A,B)\in \mathcal {AB}$, $J\subsetneq I$ and $B^J$ be such that \[
  B_i^J = \left\{
    \begin{array}{ll}
      B_i&\text{ if }i\in J\\
      \omega(i)\setminus A_i & \text{  otherwise},
    \end{array}
    \right.
  \]
  Let $\mu$ be a component-wise
  individually rational and Pareto-efficient matching  at
  $(A, B^J)$ and let $\bar B^J$ be such that \[
  \bar B_i^J = \left\{
    \begin{array}{ll}
      B_i&\text{ if }i\in J\\
      O\setminus A_i & \text{  otherwise},
    \end{array}
    \right.
  \]
Then, there is $i\in I\setminus J$ such that, for all $\mu'\in\mathcal M^0(A,\bar B^J,\mu)$, $|\mu'(i)\cap A_i|=|\mu(i)\cap A_i|$.  
  \end{lemma}

  \begin{proof}
   Assume to the contrary that $J\subsetneq I$ and, for all $i\in
   I\setminus J$, there exists $\mu^i\in\mathcal{M}^0(A,\bar B^J,\mu)$
   such that $|\mu^i(i)\cap A_i |  > |\mu(i)\cap A_i|$. Since $\mu$ is
   component-wise individually rational at $(A,B^J)$ and
   $\mu^i\in\mathcal{M}^0(A,\bar B^J,\mu)$, by lemmata \ref{prop:CIRP Pareto} and \ref{lemma: decomposition} imply that it is without
    loss of generality to suppose that there is a cycle
    $C^i$ that is component-wise individually rational at $(A,\bar B^J)$ such that $\mu^i = \mu + C^i$. Moreover, we may suppose
    that $C^i$ is of the form $(j_1^i, p_1^i, \dots, j_{M^i}^i, p_{M^i}
    ^i)$ where $j_1^i = i$, \aw{$p_1^i\in A_i$}, and $p_{M^i}^i\in \omega(i)\setminus A_i$. 
    
    We now construct a cycle that welfare-improves $\mu$ and that is component-wise individually rational at $(A, B^J)$, thus contradicting the efficiency of $\mu$ at $(A,B^J)$. 

Proceeding inductively, suppose that, for some $M\geq 1$, we have constructed a sequence of distinct agents and objects $(i_1,o_1,\dots,i_M,o_M)$ with the following properties: 
\begin{enumerate}
\item There exists an increasing sequence $(k_n)_{n=1}^N$ such that
\begin{enumerate}
\item $1=k_1<\dots<k_N\leq M$
\item $i_{k_n}\in I\setminus (J\cup\{i_{k_1},\dots,i_{k_{n-1}}\})$ for all $n$
\item $o_{k_n}\in A_{i_{k_n}}$ for all $n$ and $o_{k_n-1}\in\omega({k_{i_n}})$ for all $n\geq 2$
\item for all $m\notin\{k_1,\dots,k_N\}$ such that $i_m\in I\setminus J$, we have that $\{o_{m-1},o_{m}\}\subseteq A_{i_m}$
\end{enumerate}
\item For all $m$ such that $i_m\in J$, either $\{o_{m-1},o_m\}\subseteq A_{i_m}$ or $\{o_{m-1},o_m\}\subseteq B_{i_m}$.
\item For all $m<M$, $o_m\in\mu(i_{m+1})$.
\item For some $j\in I\setminus (J\cup\{i_{k_1},\dots,i_{k_N}\})$, $o_M\in\omega(j)\setminus A_j$.
\end{enumerate}

We show that we will either find a cycle that we seek or that we can extend the sequence to satisfy all properties above. Let $i_{M+1}$ be such that $o_M\in \mu(i_{M+1})$ and consider the cycle $C^{i_{M+1}}=(j_1^{i_{M+1}},p^{i_{M+1}}_1,\dots,j_{M^{i_{M+1}}}^{i_{M+1}},p_{M^{i_{M+1}}}^{i_{M+1}})$. 

If there does not exist an $m\in\{1,\dots,M^{i_{M+1}}\}$ such that $j^{i_{M+1}}_m\in I\setminus J$ and $p^{i_{M+1}}_m\in O\setminus A_{j^{i_{M+1}}_m}$, then $C^{i_{M+1}}$ is a cycle that we seek. 

Otherwise, let $k_{N+1}=M+1$ and let $\Delta$ be the smallest integer in $\{2,\dots,M^{i_{M+1}}\}$ such that $j_{\Delta}^{i_{M+1}}\in I\setminus J$ and $p_{\Delta}^{i_{M+1}}\in O\setminus A_{j_{\Delta}^{i_{M+1}}}$. Since $C^{i_{M+1}}$ is a cycle at $\mu$, we have that $p_{\Delta-1}^{i_{M+1}}\in \mu(j_{\Delta}^{i_{M+1}})$. Since $\mu$ is component-wise individually rational at $(A,B^J)$ and $j_{\Delta}^{i_{M+1}}\in I\setminus J$, we have that $p_{\Delta-1}^{i_{M+1}}\in \omega({j_{\Delta}^{i_{M+1}}})\setminus A_{j_{\Delta}^{i_{M+1}}}$. 

Assume first that there exists $m^*<\Delta$ such that $p_{m^*}^{i_{M+1}}=o_{m'}$ for some $m'\leq M$ and - without loss of generality - $p_{\hat m}^{i_{M+1}}\notin\{o_1,\dots,o_M\}$ for all $\hat m<m^*$. If $j_{\hat m}^{i_{M+1}}\notin \{i_{m'+1},\dots,i_M\}$ for all $\hat m\leq m^*$, then  
\[
(i_{m'+1},o_{m'+1},\dots,i_M,o_M,j_1^{i_{M+1}},p_1^{i_{M+1}},\dots,j_{m^*}^{i_{M+1}},p_{m^*}^{i_{M+1}})
\]
is a cycle that we seek. If $\hat{m}^*$ is the smallest integer $\leq \hat{m}$ such that $j_{\hat{m}^*}^{i_{M+1}}=i_{m''}$ for some $m''\geq m'+1$, then  

\[
(i_{m''},o_{m''},\dots,i_M,o_M,j_1^{i_{M+1}},p_1^{i_{M+1}},\dots,j_{\hat{m}^*-1}^{i_{M+1}},p_{\hat{m}^*-1}^{i_{M+1}})
\]
is a cycle that we seek.\footnote{Note that this also holds for the case of $\hat{m}^*=1$ since our inductive assumption implies that $o^{m''}\in A_{i^{m''}}$ and $o^M\in \omega({i^{m''}})\setminus A_{i^{m''}}$ there.}

Next, if $\{o_1,\dots,o_M\}\cap\{p_1^{i_{M+1}},\dots,p_{\Delta}^{i_{M+1}}\}=\emptyset$ but there exists some $m^*<\Delta$ such that $j_{m^*}^{i_{M+1}}=i_{m'}$ for some $m'\leq M$ and - without loss of generality - $j_{\hat m}^{i_{M+1}}\notin\{i_1,\dots,i_M\}$ for all $\hat m<m^*$, then 
\[
(i_{m'},o_{m'},\dots,i_M,o_M,j_1^{i_{M+1}},p_1^{i_{M+1}},\dots,j_{m^*-1}^{i_{M+1}},p_{m^*-1}^{i_{M+1}})
\]
is a cycle that we seek. 

In any other case, we obtain the desired extension of $(i_1,o_1,\dots,i_M,o_M)$ by setting, for all $\delta \leq \Delta-1$, $(i_{m+\delta},o_{m+\delta})=(j_{\delta}^{i_{M+1}},p_{\delta}^{i_{M+1}})$. 

Since the sets of agents and objects are both finite, we  eventually obtain a contradiction. 
\end{proof}

Before arguing how the previous two lemmas imply \cref{thm:sufficiency IR PE}, we note an interesting corollary to \cref{lemma:noimprovement}. Recall that the smallest \B set that an agent $i\in I$ can have is
$\omega(i)\setminus A_i$ and that the largest \B set is $O\setminus
A_i$. Setting, for each $i\in I$,  $\underline B_i = \omega(i)\setminus A_i$ and
$\overline B_i = O\setminus A_i$, we get the following. 

\begin{corollary}\label{cor:weak eff}
  Fix a profile of \A sets  $A$. Let $\underline B$ be the profile
  with the smallest \B sets (for each $i\in I$, $B_i =
  \omega(i)\setminus A_i$) and let $\overline B$ be the profile with
  the largest \B sets (for each $i\in I$, $B_i = O\setminus A_i$). If
  $\mu$ is unambiguously individually rational and unambiguously efficient
  at $(A, \underline B)$, then it is unambiguously  weakly efficient at $(A,  \overline B)$. 
\end{corollary}

\begin{proof}[Proof of Theorem~\ref{thm:sufficiency IR PE}]

Note first that Lemma \ref{lemma:noimprovement} immediately implies that our algorithm above is well
defined. Now given a profile $(A,B)\in\mathcal{AB}$, let $(I^t)_{t=1}^T$
be the sequence of sets of agents, who can no longer improve (recall
that $I^T=I$), $(\mu^t)_{t=1}^T$ be a sequence of matchings such that,
for all $t$, $\mu^t\in\mathcal{M}^n(A,B^{t-1},\mu^{t-1})$, and
$\mathcal{M}^n(A,B^{T},\mu^T)$ be the set of
matchings that remain at the end of our algorithm. By the construction
of $\mathcal{M}^n(A,B^{T},\mu^T)$, there cannot be a component-wise individually
rational and Pareto-improving cycle of any matching it contains. By
Lemma~\ref{prop:CIRP Pareto}, we immediately obtain that all matchings
in $\mathcal{M}^n(A,B^{T},\mu^T)$ are  unambiguously efficient. 

\end{proof}

\bibliographystyle{ACM-Reference-Format}
\bibliography{refs}


\begin{thebibliography}{36}


\ifx \showCODEN    \undefined \def \showCODEN     #1{\unskip}     \fi
\ifx \showDOI      \undefined \def \showDOI       #1{#1}\fi
\ifx \showISBNx    \undefined \def \showISBNx     #1{\unskip}     \fi
\ifx \showISBNxiii \undefined \def \showISBNxiii  #1{\unskip}     \fi
\ifx \showISSN     \undefined \def \showISSN      #1{\unskip}     \fi
\ifx \showLCCN     \undefined \def \showLCCN      #1{\unskip}     \fi
\ifx \shownote     \undefined \def \shownote      #1{#1}          \fi
\ifx \showarticletitle \undefined \def \showarticletitle #1{#1}   \fi
\ifx \showURL      \undefined \def \showURL       {\relax}        \fi
\providecommand\bibfield[2]{#2}
\providecommand\bibinfo[2]{#2}
\providecommand\natexlab[1]{#1}
\providecommand\showeprint[2][]{arXiv:#2}

\bibitem[\protect\citeauthoryear{Alcalde-Unzu and Molis}{Alcalde-Unzu and
  Molis}{2011}]%
        {AlcaldeMolis:GEB2011}
\bibfield{author}{\bibinfo{person}{Jorge Alcalde-Unzu} {and}
  \bibinfo{person}{Elena Molis}.} \bibinfo{year}{2011}\natexlab{}.
\newblock \showarticletitle{Exchange of indivisible goods and indifferences:
  The Top Trading Absorbing Sets mechanisms}.
\newblock \bibinfo{journal}{\emph{Games and Economic Behavior}}
  \bibinfo{volume}{73}, \bibinfo{number}{1} (\bibinfo{date}{September}
  \bibinfo{year}{2011}), \bibinfo{pages}{1--16}.
\newblock


\bibitem[\protect\citeauthoryear{Alwin}{Alwin}{2007}]%
        {Alwin2007}
\bibfield{author}{\bibinfo{person}{Duane~F Alwin}.}
  \bibinfo{year}{2007}\natexlab{}.
\newblock \bibinfo{booktitle}{\emph{Margins of error: A study of reliability in
  survey measurement}}.
\newblock \bibinfo{publisher}{John Wiley \& Sons}.
\newblock


\bibitem[\protect\citeauthoryear{Andersson, \'{A}gnes Cseh, Ehlers, and
  Erlanson}{Andersson et~al\mbox{.}}{2021}]%
        {AnderssonCsehEhlersErlanson:2018}
\bibfield{author}{\bibinfo{person}{Tommy Andersson}, \bibinfo{person}{\'{A}gnes
  Cseh}, \bibinfo{person}{Lars Ehlers}, {and} \bibinfo{person}{Albin
  Erlanson}.} \bibinfo{year}{2021}\natexlab{}.
\newblock \showarticletitle{Organizing Time Banks: Lessons from Matching
  Markets}.
\newblock \bibinfo{journal}{\emph{American Economic Journal:Microeconomics}}
  \bibinfo{volume}{13}, \bibinfo{number}{1} (\bibinfo{date}{February}
  \bibinfo{year}{2021}), \bibinfo{pages}{338--73}.
\newblock


\bibitem[\protect\citeauthoryear{Aziz, Biró, Lang, Lesca, and Monnot}{Aziz
  et~al\mbox{.}}{2019}]%
        {AzizEtAl2019}
\bibfield{author}{\bibinfo{person}{Haris Aziz}, \bibinfo{person}{Péter Biró},
  \bibinfo{person}{Jérôme Lang}, \bibinfo{person}{Julien Lesca}, {and}
  \bibinfo{person}{Jérôme Monnot}.} \bibinfo{year}{2019}\natexlab{}.
\newblock \showarticletitle{Efficient reallocation under additive and
  responsive preferences}.
\newblock \bibinfo{journal}{\emph{Theoretical Computer Science}}
  \bibinfo{volume}{790} (\bibinfo{year}{2019}), \bibinfo{pages}{1--15}.
\newblock


\bibitem[\protect\citeauthoryear{Aziz, Islam, and Pápai}{Aziz
  et~al\mbox{.}}{2025}]%
        {aziz2025strategyproofmaximummatchingdichotomous}
\bibfield{author}{\bibinfo{person}{Haris Aziz}, \bibinfo{person}{Md.~Shahidul
  Islam}, {and} \bibinfo{person}{Szilvia Pápai}.}
  \bibinfo{year}{2025}\natexlab{}.
\newblock \bibinfo{title}{Strategyproof Maximum Matching under Dichotomous
  Agent Preferences}.
\newblock
\newblock
\showeprint[arxiv]{2502.09962}~[cs.GT]
\urldef\tempurl%
\url{https://arxiv.org/abs/2502.09962}
\showURL{%
\tempurl}


\bibitem[\protect\citeauthoryear{Biró, Klijn, and Pápai}{Biró
  et~al\mbox{.}}{2022}]%
        {BiroKlijnPapai:2018}
\bibfield{author}{\bibinfo{person}{Péter Biró}, \bibinfo{person}{Flip Klijn},
  {and} \bibinfo{person}{Szilvia Pápai}.} \bibinfo{year}{2022}\natexlab{}.
\newblock \showarticletitle{Serial Rules in a Multi-Unit Shapley-Scarf Market}.
\newblock \bibinfo{journal}{\emph{Games and Economic Behavior}}
  \bibinfo{volume}{136} (\bibinfo{year}{2022}), \bibinfo{pages}{428--453}.
\newblock
\showISSN{0899-8256}


\bibitem[\protect\citeauthoryear{Bogomolnaia and Moulin}{Bogomolnaia and
  Moulin}{2004}]%
        {BogomolnaiaMoulin:Econometrica2004}
\bibfield{author}{\bibinfo{person}{Anna Bogomolnaia} {and}
  \bibinfo{person}{Herv\'{e} Moulin}.} \bibinfo{year}{2004}\natexlab{}.
\newblock \showarticletitle{Random Matching Under Dichotomous Preferences}.
\newblock \bibinfo{journal}{\emph{Econometrica}} \bibinfo{volume}{72},
  \bibinfo{number}{1} (\bibinfo{date}{01} \bibinfo{year}{2004}),
  \bibinfo{pages}{257--279}.
\newblock


\bibitem[\protect\citeauthoryear{Bouveret, Endriss, and Lang}{Bouveret
  et~al\mbox{.}}{2010}]%
        {BouveretEndrissLang2010}
\bibfield{author}{\bibinfo{person}{Sylvain Bouveret}, \bibinfo{person}{Ulle
  Endriss}, {and} \bibinfo{person}{J\'{e}r\^{o}me Lang}.}
  \bibinfo{year}{2010}\natexlab{}.
\newblock \showarticletitle{Fair Division under Ordinal Preferences: Computing
  Envy-Free Allocations of Indivisible Goods}. In
  \bibinfo{booktitle}{\emph{Proceedings of the 2010 Conference on ECAI 2010:
  19th European Conference on Artificial Intelligence}}.
  \bibinfo{publisher}{IOS Press}, \bibinfo{address}{NLD},
  \bibinfo{pages}{387–392}.
\newblock


\bibitem[\protect\citeauthoryear{Brams, Edelman, and Fishburn}{Brams
  et~al\mbox{.}}{2003}]%
        {BramsEdelmanFishburnTD2003}
\bibfield{author}{\bibinfo{person}{Steven Brams}, \bibinfo{person}{Paul
  Edelman}, {and} \bibinfo{person}{Peter Fishburn}.}
  \bibinfo{year}{2003}\natexlab{}.
\newblock \showarticletitle{Fair division of indivisible items}.
\newblock \bibinfo{journal}{\emph{Theory and Decision}}  \bibinfo{volume}{55}
  (\bibinfo{date}{02} \bibinfo{year}{2003}), \bibinfo{pages}{147--180}.
\newblock


\bibitem[\protect\citeauthoryear{Budish and Kessler}{Budish and
  Kessler}{2018}]%
        {BudishKessler:2018}
\bibfield{author}{\bibinfo{person}{Eric Budish} {and} \bibinfo{person}{Judd~B.
  Kessler}.} \bibinfo{year}{2018}\natexlab{}.
\newblock \bibinfo{booktitle}{\emph{Bringing Real Market Participants' Real
  Preferences into the Lab: An Experiment that Changed the Course Allocation
  Mechanism at Wharton}}.
\newblock \bibinfo{type}{NBER Working Papers} 22448.
  \bibinfo{institution}{National Bureau of Economic Research, Inc}.
\newblock


\bibitem[\protect\citeauthoryear{Chernev}{Chernev}{2003}]%
        {Chernev:JCR2003}
\bibfield{author}{\bibinfo{person}{Alexander Chernev}.}
  \bibinfo{year}{2003}\natexlab{}.
\newblock \showarticletitle{When More Is Less and Less Is More: The Role of
  Ideal Point Availability and Assortment in Consumer Choice}.
\newblock \bibinfo{journal}{\emph{Journal of Consumer Research}}
  \bibinfo{volume}{30}, \bibinfo{number}{2} (\bibinfo{year}{2003}),
  \bibinfo{pages}{170--183}.
\newblock


\bibitem[\protect\citeauthoryear{Dhar}{Dhar}{1997}]%
        {Dhar:JCR1997}
\bibfield{author}{\bibinfo{person}{Ravi Dhar}.}
  \bibinfo{year}{1997}\natexlab{}.
\newblock \showarticletitle{Consumer Preference for a No‐Choice Option}.
\newblock \bibinfo{journal}{\emph{Journal of Consumer Research}}
  \bibinfo{volume}{25}, \bibinfo{number}{2} (\bibinfo{year}{1997}),
  \bibinfo{pages}{215--231}.
\newblock


\bibitem[\protect\citeauthoryear{Ehlers}{Ehlers}{2008}]%
        {ehlers2008}
\bibfield{author}{\bibinfo{person}{Lars Ehlers}.}
  \bibinfo{year}{2008}\natexlab{}.
\newblock \showarticletitle{Truncation strategies in matching markets}.
\newblock \bibinfo{journal}{\emph{Mathematics of Operations Research}}
  \bibinfo{volume}{33}, \bibinfo{number}{2} (\bibinfo{year}{2008}).
\newblock


\bibitem[\protect\citeauthoryear{Ehlers and Klaus}{Ehlers and Klaus}{2003}]%
        {EhlersKlausSCW2003}
\bibfield{author}{\bibinfo{person}{Lars Ehlers} {and} \bibinfo{person}{Bettina
  Klaus}.} \bibinfo{year}{2003}\natexlab{}.
\newblock \showarticletitle{Coalitional strategy-proof and resource-monotonic
  solutions for multiple assignment problems}.
\newblock \bibinfo{journal}{\emph{Social Choice and Welfare}}
  \bibinfo{volume}{21} (\bibinfo{date}{02} \bibinfo{year}{2003}),
  \bibinfo{pages}{265--280}.
\newblock


\bibitem[\protect\citeauthoryear{Ergin, S\"{o}nmez, and \"{U}nver}{Ergin
  et~al\mbox{.}}{2020}]%
        {Erginetal2020}
\bibfield{author}{\bibinfo{person}{Haluk Ergin}, \bibinfo{person}{Tayfun
  S\"{o}nmez}, {and} \bibinfo{person}{M.~Utku \"{U}nver}.}
  \bibinfo{year}{2020}\natexlab{}.
\newblock \showarticletitle{Efficient and incentive-compatible liver exchange}.
\newblock \bibinfo{journal}{\emph{Econometrica}} \bibinfo{volume}{88},
  \bibinfo{number}{3} (\bibinfo{date}{May} \bibinfo{year}{2020}),
  \bibinfo{pages}{1645 -- 1671}.
\newblock


\bibitem[\protect\citeauthoryear{Greenleaf and Lehmann}{Greenleaf and
  Lehmann}{1995}]%
        {GreenleafLehmann:JCR1995}
\bibfield{author}{\bibinfo{person}{Eric~A. Greenleaf} {and}
  \bibinfo{person}{Donald~R. Lehmann}.} \bibinfo{year}{1995}\natexlab{}.
\newblock \showarticletitle{Reasons for Substantial Delay in Consumer Decision
  Making}.
\newblock \bibinfo{journal}{\emph{Journal of Consumer Research}}
  \bibinfo{volume}{22}, \bibinfo{number}{2} (\bibinfo{year}{1995}),
  \bibinfo{pages}{186--199}.
\newblock


\bibitem[\protect\citeauthoryear{Iyengar and Lepper}{Iyengar and
  Lepper}{2000}]%
        {IyengarLepper:2000JPSP}
\bibfield{author}{\bibinfo{person}{Sheena~S. Iyengar} {and}
  \bibinfo{person}{Mark~R. Lepper}.} \bibinfo{year}{2000}\natexlab{}.
\newblock \showarticletitle{When Choice is Demotivating: Can One Desire Too
  Much of a Good Thing?}
\newblock \bibinfo{journal}{\emph{Journal of Personality and Social
  Psychology}} \bibinfo{volume}{79}, \bibinfo{number}{6}
  (\bibinfo{year}{2000}), \bibinfo{pages}{995--1006}.
\newblock


\bibitem[\protect\citeauthoryear{Jaramillo and Manjunath}{Jaramillo and
  Manjunath}{2012}]%
        {JaramilloManjunath:JET2012}
\bibfield{author}{\bibinfo{person}{Paula Jaramillo} {and}
  \bibinfo{person}{Vikram Manjunath}.} \bibinfo{year}{2012}\natexlab{}.
\newblock \showarticletitle{The difference indifference makes in strategy-proof
  allocation of objects}.
\newblock \bibinfo{journal}{\emph{Journal of Economic Theory}}
  \bibinfo{volume}{147}, \bibinfo{number}{5} (\bibinfo{year}{2012}),
  \bibinfo{pages}{1913--1946}.
\newblock


\bibitem[\protect\citeauthoryear{Konishi, Quint, and Wako}{Konishi
  et~al\mbox{.}}{2001}]%
        {KonishiQuintWako:JME2001}
\bibfield{author}{\bibinfo{person}{Hideo Konishi}, \bibinfo{person}{Thomas
  Quint}, {and} \bibinfo{person}{Jun Wako}.} \bibinfo{year}{2001}\natexlab{}.
\newblock \showarticletitle{On the {S}hapley--{S}carf economy: The case of
  multiple types of indivisible goods}.
\newblock \bibinfo{journal}{\emph{Journal of Mathematical Economics}}
  \bibinfo{volume}{35}, \bibinfo{number}{1} (\bibinfo{year}{2001}),
  \bibinfo{pages}{1 -- 15}.
\newblock
\showISSN{0304-4068}


\bibitem[\protect\citeauthoryear{Manjunath and Westkamp}{Manjunath and
  Westkamp}{2021}]%
        {ManjunathWestkamp:2018a}
\bibfield{author}{\bibinfo{person}{Vikram Manjunath} {and}
  \bibinfo{person}{Alexander Westkamp}.} \bibinfo{year}{2021}\natexlab{}.
\newblock \showarticletitle{Strategy-Proof Exchange under Trichotomous
  Preferences}.
\newblock \bibinfo{journal}{\emph{Journal of Economic Theory}}
  \bibinfo{volume}{193} (\bibinfo{year}{2021}), \bibinfo{pages}{105197}.
\newblock


\bibitem[\protect\citeauthoryear{Milgrom}{Milgrom}{2009}]%
        {Milgrom:AEJmicro2009}
\bibfield{author}{\bibinfo{person}{Paul Milgrom}.}
  \bibinfo{year}{2009}\natexlab{}.
\newblock \showarticletitle{Assignment Messages and Exchanges}.
\newblock \bibinfo{journal}{\emph{American Economic Journal: Microeconomics}}
  \bibinfo{volume}{1}, \bibinfo{number}{2} (\bibinfo{date}{August}
  \bibinfo{year}{2009}), \bibinfo{pages}{95--113}.
\newblock


\bibitem[\protect\citeauthoryear{Milgrom}{Milgrom}{2010}]%
        {Milgrom:GEB2010}
\bibfield{author}{\bibinfo{person}{Paul Milgrom}.}
  \bibinfo{year}{2010}\natexlab{}.
\newblock \showarticletitle{Simplified mechanisms with an application to
  sponsored-search auctions}.
\newblock \bibinfo{journal}{\emph{Games and Economic Behavior}}
  \bibinfo{volume}{70}, \bibinfo{number}{1} (\bibinfo{date}{September}
  \bibinfo{year}{2010}), \bibinfo{pages}{62--70}.
\newblock


\bibitem[\protect\citeauthoryear{Milgrom}{Milgrom}{2011}]%
        {Milgrom:EI2011}
\bibfield{author}{\bibinfo{person}{Paul Milgrom}.}
  \bibinfo{year}{2011}\natexlab{}.
\newblock \showarticletitle{Critical issues in the practice of market design}.
\newblock \bibinfo{journal}{\emph{Economic Inquiry}} \bibinfo{volume}{49},
  \bibinfo{number}{2} (\bibinfo{year}{2011}), \bibinfo{pages}{311--320}.
\newblock


\bibitem[\protect\citeauthoryear{Ortega}{Ortega}{2020}]%
        {Ortega2020}
\bibfield{author}{\bibinfo{person}{Josu{\'e} Ortega}.}
  \bibinfo{year}{2020}\natexlab{}.
\newblock \showarticletitle{Multi-unit assignment under dichotomous
  preferences}.
\newblock \bibinfo{journal}{\emph{Mathematical Social Sciences}}
  \bibinfo{volume}{103} (\bibinfo{year}{2020}), \bibinfo{pages}{15--24}.
\newblock
\urldef\tempurl%
\url{https://doi.org/10.1016/j.mathsocsci.2019.11.003}
\showDOI{\tempurl}


\bibitem[\protect\citeauthoryear{Othman, Sandholm, and Budish}{Othman
  et~al\mbox{.}}{2010}]%
        {OthmanSandholmBudish:AAMAS2010}
\bibfield{author}{\bibinfo{person}{Abraham Othman}, \bibinfo{person}{Tuomas
  Sandholm}, {and} \bibinfo{person}{Eric Budish}.}
  \bibinfo{year}{2010}\natexlab{}.
\newblock \showarticletitle{Finding Approximate Competitive Equilibria:
  Efficient and Fair Course Allocation}. In
  \bibinfo{booktitle}{\emph{Proceedings of the 9th International Conference on
  Autonomous Agents and Multiagent Systems: Volume 1 - Volume 1}}
  \emph{(\bibinfo{series}{AAMAS '10})}. \bibinfo{publisher}{International
  Foundation for Autonomous Agents and Multiagent Systems},
  \bibinfo{pages}{873--880}.
\newblock
\showISBNx{9780982657119}


\bibitem[\protect\citeauthoryear{P\`{a}pai}{P\`{a}pai}{2001}]%
        {PapaiJET2001}
\bibfield{author}{\bibinfo{person}{Szilvia P\`{a}pai}.}
  \bibinfo{year}{2001}\natexlab{}.
\newblock \showarticletitle{Strategyproof and Nonbossy Multiple Assignments}.
\newblock \bibinfo{journal}{\emph{Journal of Public Economic Theory}}
  \bibinfo{volume}{3} (\bibinfo{date}{07} \bibinfo{year}{2001}),
  \bibinfo{pages}{257--71}.
\newblock


\bibitem[\protect\citeauthoryear{P\`{a}pai}{P\`{a}pai}{2003}]%
        {Papai:JME2003}
\bibfield{author}{\bibinfo{person}{Szilvia P\`{a}pai}.}
  \bibinfo{year}{2003}\natexlab{}.
\newblock \showarticletitle{Strategyproof exchange of indivisible goods}.
\newblock \bibinfo{journal}{\emph{Journal of Mathematical Economics}}
  \bibinfo{volume}{39}, \bibinfo{number}{8} (\bibinfo{date}{November}
  \bibinfo{year}{2003}), \bibinfo{pages}{931--959}.
\newblock


\bibitem[\protect\citeauthoryear{Roth}{Roth}{1985}]%
        {Roth1985}
\bibfield{author}{\bibinfo{person}{Al Roth}.} \bibinfo{year}{1985}\natexlab{}.
\newblock \showarticletitle{The college admissions problem is not equivalent to
  the marriage problem}.
\newblock \bibinfo{journal}{\emph{Journal of Economic Theory}}
  \bibinfo{volume}{36} (\bibinfo{year}{1985}), \bibinfo{pages}{277--288}.
\newblock


\bibitem[\protect\citeauthoryear{Roth}{Roth}{1982}]%
        {Roth:EL1982}
\bibfield{author}{\bibinfo{person}{Alvin~E. Roth}.}
  \bibinfo{year}{1982}\natexlab{}.
\newblock \showarticletitle{Incentive compatibility in a market with
  indivisible goods}.
\newblock \bibinfo{journal}{\emph{Economics Letters}} \bibinfo{volume}{9},
  \bibinfo{number}{2} (\bibinfo{year}{1982}), \bibinfo{pages}{127--132}.
\newblock


\bibitem[\protect\citeauthoryear{Roth and Postlewaite}{Roth and
  Postlewaite}{1977}]%
        {RothPostlewaite:JME1977}
\bibfield{author}{\bibinfo{person}{Alvin~E. Roth} {and} \bibinfo{person}{Andrew
  Postlewaite}.} \bibinfo{year}{1977}\natexlab{}.
\newblock \showarticletitle{Weak versus strong domination in a market with
  indivisible goods}.
\newblock \bibinfo{journal}{\emph{Journal of Mathematical Economics}}
  \bibinfo{volume}{4}, \bibinfo{number}{2} (\bibinfo{date}{August}
  \bibinfo{year}{1977}), \bibinfo{pages}{131--137}.
\newblock


\bibitem[\protect\citeauthoryear{Roth and Rothblum}{Roth and Rothblum}{1999}]%
        {rothrothblum1999}
\bibfield{author}{\bibinfo{person}{Alvin~E. Roth} {and}
  \bibinfo{person}{Uriel~G. Rothblum}.} \bibinfo{year}{1999}\natexlab{}.
\newblock \showarticletitle{Truncation strategies in matching markets - In
  search of advice for participants}.
\newblock \bibinfo{journal}{\emph{Econometrica}}  \bibinfo{volume}{67}
  (\bibinfo{year}{1999}), \bibinfo{pages}{21 -- 43}.
\newblock


\bibitem[\protect\citeauthoryear{Shapley and Scarf}{Shapley and Scarf}{1974}]%
        {ShapleyScarf:JME1974}
\bibfield{author}{\bibinfo{person}{Lloyd Shapley} {and}
  \bibinfo{person}{Herbert Scarf}.} \bibinfo{year}{1974}\natexlab{}.
\newblock \showarticletitle{On cores and indivisibility}.
\newblock \bibinfo{journal}{\emph{Journal of Mathematical Economics}}
  \bibinfo{volume}{1} (\bibinfo{year}{1974}), \bibinfo{pages}{23--37}.
\newblock


\bibitem[\protect\citeauthoryear{S\"onmez}{S\"onmez}{1999}]%
        {Sonmez:1999ecta}
\bibfield{author}{\bibinfo{person}{Tayfun S\"onmez}.}
  \bibinfo{year}{1999}\natexlab{}.
\newblock \showarticletitle{Strategy-Proofness and Essentially Single-Valued
  Cores}.
\newblock \bibinfo{journal}{\emph{Econometrica}} \bibinfo{volume}{67},
  \bibinfo{number}{3} (\bibinfo{date}{May} \bibinfo{year}{1999}),
  \bibinfo{pages}{677--690}.
\newblock


\bibitem[\protect\citeauthoryear{S\"onmez and Yenmez}{S\"onmez and
  Yenmez}{2020}]%
        {sonmezyenmez20}
\bibfield{author}{\bibinfo{person}{Tayfun S\"onmez} {and}
  \bibinfo{person}{Bumin Yenmez}.} \bibinfo{year}{2020}\natexlab{}.
\newblock \showarticletitle{Incentivized kidney exchange}.
\newblock \bibinfo{journal}{\emph{American Economic Review}}
  \bibinfo{volume}{110} (\bibinfo{year}{2020}), \bibinfo{pages}{2198 -- 2224}.
\newblock


\bibitem[\protect\citeauthoryear{Todo, Sun, and Yokoo}{Todo
  et~al\mbox{.}}{2014}]%
        {TodoSunYokoo2014}
\bibfield{author}{\bibinfo{person}{Taiki Todo}, \bibinfo{person}{Haixin Sun},
  {and} \bibinfo{person}{Makoto Yokoo}.} \bibinfo{year}{2014}\natexlab{}.
\newblock \showarticletitle{Strategyproof Exchange with Multiple Private
  Endowments}.
\newblock \bibinfo{journal}{\emph{Proceedings of the National Conference on
  Artificial Intelligence}}  \bibinfo{volume}{1} (\bibinfo{date}{06}
  \bibinfo{year}{2014}), \bibinfo{pages}{805--811}.
\newblock


\bibitem[\protect\citeauthoryear{Troyan and Morrill}{Troyan and
  Morrill}{2020}]%
        {troyanmorrill}
\bibfield{author}{\bibinfo{person}{Peter Troyan} {and} \bibinfo{person}{Thayer
  Morrill}.} \bibinfo{year}{2020}\natexlab{}.
\newblock \showarticletitle{Obvious manipulations}.
\newblock \bibinfo{journal}{\emph{Journal of Economic Theory}}
  \bibinfo{volume}{185} (\bibinfo{year}{2020}), \bibinfo{pages}{104970}.
\newblock


\end{thebibliography}

\newpage

\appendix

\crefalias{section}{appendix}

\section{Proof of \cref{prop: weak dom truth}}
\label{proof of weak dom truth}

Fix a finite set of agents $I=\{1,\dots,n\}$ and a finite set of objects $O$. Fix a profile of trichotomous preferences $(A,B)$ and let $\{\mathcal M^t\}_{t=1}^n$ be the sequence of sets of allocations computed via the IRP procedure. Fix an arbitrary agent $i^*\in I$.  

Consider some $o^*\in O\setminus (A_{i^*}\cup B_{i^*})$. Let
{$(\hat{A},\hat{B})$ be such that} 
$o^*\in \hat{A}_{i^*}\cup
\hat{B}_{i^*}${, $\hat A_{i^*}\setminus\{o^*\} = A_{i^*}$, $\hat B_{i^*}\setminus\{o^*\} = B_{i^*}$, and}
$(\hat{A}_j,\hat{B}_j)=(A_j,B_j)$ for all $j\neq i^*$. Let
$\{\hat{\mathcal M}^t\}_{t=1}^n$ be the sequence of sets of
allocations computed via the IRP procedure for the new input
$(\hat{A},\hat{B})$. Let $\hat{\mu}\in\hat{\mathcal{M}}^n$ be
arbitrary.  

\begin{claim}
\begin{enumerate}
\item If $o^*\in \hat{A}_{i^*}$, then $$0\leq |\hat{\mu}(i^*)\cap \hat{A}_{i^*}|-|\mu(i^*)\cap A_{i^*}|\leq 1$$ and $o^*\in \hat{\mu}(i^*)$ if $|\hat{\mu}(i^*)\cap \hat{A}_{i^*}|-|\mu(i^*)\cap A_{i^*}|= 1$.
\item If $o^*\in \hat{B}_{i^*}$, then { $$0\leq |\mu(i)\cap
  A_{i^*}|-|\hat{\mu}(i^*)\cap A_{i^*}|\leq 1$$} and $o^*\in \hat{\mu}(i^*)$ if {$ |\mu(i)\cap A_{i^*}|-|\hat{\mu}(i^*)\cap A_{i^*}|=1$.}
\end{enumerate}
\end{claim}

\begin{proof}

Define a bipartite graph $G$ that has, for each $i\in I$, two vertices $a(i)$ and $b(i)$, and on the right side one vertex for each object in $O$. For each $i$ and each object $o\in \hat{A}_i$, there is edge connecting $a(i)$ and $o$, and for each object $p\in \hat{B}_i$, there is an edge connecting $b(i)$ and $p$.

Next, we introduce a two-colored subgraph $H$ of $G$ that plays an important role in our proof: 
\begin{itemize}
\item If $o\in (\mu(i)\cap \hat{A}_i)\setminus \hat{\mu}(i)$, add a \emph{red} edge connecting $a(i)$ and $o$
\item If $o\in (\mu(i)\cap \hat{B}_i)\setminus \hat{\mu}(i)$, add a \emph{red} edge connecting $b(i)$ and $o$
\item If $o\in (\hat{\mu}(i)\cap \hat{A}_i)\setminus \mu(i)$, add a \emph{blue} edge connecting $a(i)$ and $o$
\item If $o\in (\hat{\mu}(i)\cap \hat{B}_i)\setminus \mu(i)$, add a \emph{blue} edge connecting $b(i)$ and $o$
\end{itemize}

For the remainder of the proof, we assume that there is at least one
agent $j$ for whom $|\mu(j)\cap \hat{A}_j|\neq |\hat{\mu(j)}\cap A_j|$
and let $j^*$ be the lowest indexed (highest priority) such agent.  

Let $C^1,\dots,C^K$ be a collection of {alternating (red and blue)} cycles of $H$ such that when we
remove the edges associated with these cycles, we obtain an acyclic
subgraph $H^*$ of $H$. For any $k$, let $\mu^k$ be defined by, for
each agent $i$,  
\begin{itemize}
\item \emph{removing} all objects $o\in\mu^{k-1}(i)$ that are associated with a red edge in $C^k$ that is incident to some $i$-copy in $H$ and 
\item \emph{adding} all objects $o$ to $\mu^{k-1}(i)$ that are associated with a blue edge in $C^k$ that is incident to some $i$-copy in $H$,
\end{itemize}
where $\mu^0\equiv \mu$. By construction of $H$ {since the  $C^k$s
  contain alternating red and blue edges}, we have that $|\mu^k(j)\cap \hat{A}_j|=|\mu(j)\cap A_j|$ for all $k$ and all $j$. 

If there are no remaining edges in $H^*$, the preceding observation immediately implies that we get a contradiction to our assumption that $|\mu(j^*)\cap A_{j^*}|\neq |\hat{\mu}(j^*)\cap \hat{A}_{j^*}|$. Henceforth, we thus assume that $H^*$ is non-empty. Since $H^*$ is acyclic and non-empty, there must exist an agent copy $c$ that is only incident to red edges in $H^*$ and another agent copy $c'$ (potentially of a different agent) that is only incident to blue edges in $H^*$: If each agent copy with at least one remaining edge in $H^*$ is incident to at least one red and at least one blue edge, there must be a cycle; if there is an agent copy $c$ that is only incident to blue edges but no agent copy that is only incident to red edges, then a path that starts at $c$ and alternates between red and blue colors, will never end since there is no agent with only red edges.

Now let $c^*$ be an agent copy that is only incident to red edges. We argue that there has to exist a path in $H^*$ that connects $c^*$ to an agent copy $\hat{c}^*$ that is only incident to blue edges. Pick an arbitrary red edge $r^1$ associated with $c^*$ and let $o^1$ be the object incident to that edge. By construction, there has to be a blue edge $b^1$ incident to $o^1$ in $H^*$. Let $c^1$ be the agent copy incident to the $b^1$-edge. If $c^1$ has no red edges, we have our desired path. Otherwise, $c^1$ has a red edge $r^2$ connecting it to some object $o^2$. By construction, there has to be a blue edge $b^2$ incident to $o^2$ in $H^*$. Let $c^2$ be the agent copy incident to the $b^2$-edge. Since $H^*$ has no directed cycles, $c^2\notin\{c^*,c^1\}$. If $c^2$ has no red edges, we are done. Otherwise, we continue in the same fashion. Clearly, the process just sketched can only come to an end once we encounter an agent copy $\hat{c}^*$ that only has red edges in $H^*$, which is exactly what we wanted to show.

Let the path whose existence we just established be denoted by $P^1$,
let $\mathbf{red}^1$ be the agent copy that is only incident to one
red edge in $P^1$, and $\mathbf{blue}^1$ be the agent copy that is
only incident to one blue edge in {$P^1$}. Remove all edges in $P^1$
from $H^*$. Let the resulting subgraph of $H^*$ be denoted by
$H^1$. Continuing in the just described fashion, we get a
decomposition of the edges of $H^1$ into a finite collection of paths
$P^1,\dots,P^Z$ with the property that, for each $z$, $P^z$ connects
one agent copy $\mathbf{red}^z$ with only red edges in $H^{z-1}$ to an
agent copy $\mathbf{blue}^z$ with only blue edges in $H^{z-1}$, where
$H^0\equiv H^*$.  

Now let $i\in I$ be any agent, for whom $m_i\equiv |\mu(i)\cap A_i|\neq |\hat{\mu}(i)\cap \hat{A}_i|=\hat{m}_i$ and let $l_i=|m_i-\hat{m}_i|$. Assume first that $\hat{m}_i>m_i$. We claim that the following properties hold:
\begin{enumerate}
\item There does not exist an index $z$ such that $\mathbf{red}^z=a(i)$ or $\mathbf{blue}^z=b(i)$. 
\item There exist exactly $l_i$ indices $z_1,\dots,z_{l_i}$ such that, for each $l\in\{1,\dots,l_i\}$, $\mathbf{red}^{z_l}=b(i)$. 
\item There exist exactly $l_i$ indices $z^1,\dots,z^{l_i}$ such that, for each $l\in\{1,\dots,l_i\}$, $\mathbf{blue}^{z^l}=a(i)$.
\end{enumerate}
Let $z^*$ be the first index such that either $\mathbf{red}^{z^*}$ or
$\mathbf{blue}^{z^*}$ is an $i$ copy. If $\mathbf{red}^{z^*}=a(i)$,
our construction ensures that in $H^{z^*-1}$ and all of its subgraphs,
$a(i)$ only has red edges. By balancedness, it has to be that $b(i)$
copy has exactly as many excess blue edges as $a(i)$ has excess red
edges in $H^{z^*-1}$. Finally, by the definition of $z^*$, agent $i$
is not affected by the exchanges represented by
$P^1,\dots,P^{z^*-1}$. Summing up, we obtain a contradiction to our
assumption that $\hat{m}_i>m_i$ {because every path that $i$ is
  involved in as an endpoint has a red edge at $a(i)$ and a blue edge
  at $b(i)$}. The argument in case of
$\mathbf{blue}^{z^*}=b(i)$ is completely analogous. Hence, we must
have $\mathbf{red}^{z^*}=b(i)$ or $\mathbf{blue}^{z^*}=a(i)$. Given
balancedness, we then almost immediately obtain the first
statement. For the second and third statement, note that
$P^1,\dots,P^Z$ partition $H^*$ and an agent $i$ can only increase her
holdings of $\hat{A}$-objects by one unit when $\mathbf{blue}^z=a(i)$
for some $z$. By balancedness, there have to be exactly as many paths
$z'$ for which $\mathbf{red}^{z'}=b(i)$. 

If $m_i>\hat{m}_i$, a completely analogous argument establishes that the following properties hold:
\begin{enumerate}
\item There does not exist an index $z$ such that $\mathbf{red}^z=b(i)$ or $\mathbf{blue}^z=a(i)$. 
\item There exist exactly $l_i$ indices $z_1,\dots,z_{l_i}$ such that, for each $l\in\{1,\dots,l_i\}$, $\mathbf{red}^{z_l}=a(i)$. 
\item There exist exactly $l_i$ indices $z^1,\dots,z^{l_i}$ such that, for each $l\in\{1,\dots,l_i\}$, $\mathbf{blue}^{z^l}=b(i)$.
\end{enumerate}

Now call a path $P^z$ \emph{improving} for agent $i$, if
$\mathbf{blue}^z=a(i)$, and \emph{worsening} for agent $i$, if
$\mathbf{blue}^z=b(i)$.
{This welfare effect   is with respect to $(\hat A_i, \hat B_i)$
  which equals $(A_i, B_i)$ for all but $i=i^*$.}

By the above claim, for any agent $i$,  
\begin{itemize}
\item exactly $l_i$ of the $Z$ paths in $H^*$ are improving and there is no worsening path in $H^*$ if $\hat{m}_i>m_i$ and
\item exactly $l_i$ of the $Z$ paths in $H^*$ are worsening and there is no improving path  in $H^*$ if $m_i>\hat{m}_i$.
\end{itemize}

Next, we introduce one final auxiliary concept: Define a directed
multi-graph $D$ on node set $I$ by adding an edge from agent $i$ to
agent $j$, if there is a $z$ such that
$\mathbf{red}^z\in\{a(i),b(i)\}$ and
$\mathbf{blue}^z\in\{a(j),b(j)\}$. With a slight abuse of notation, we
use $z$ and $(i,j)$ interchangeably. By balancedness,   every node in  $D$ has the same in and out degree, so it
is Eulerian. Therefore,  we can decompose the set of edges into an edge-disjoint set of cycles where each vertex appears only once.

We are now ready to complete our proof. Assume first that there is no blue edge connecting $i^*$ and $o^*$ in $H^*$. By our assumption that $j^*$ is not indifferent between $\mu$ and $\hat{\mu}$, there has to be a directed cycle $C$ in $D$ that contains $j^*$. By definition of $j^*$, the cycle cannot contain any agent $j<j^*$. By what we have shown above, $C$ will either make $j^*$ better off or worse off. We treat these two cases separately:

\begin{itemize}
\item[\textbf{Case 1}] $C$ makes $j^*$ worse off

As $A_j\subseteq \hat{A}_j$ and $B_j\subseteq \hat{B}_j$ for all $j$,
$\mu$ is CIR at $(\hat A,\hat B)$. Since $C$ - and all other cycles in $D$ - makes $j^*$ worse off and there is no agent with higher priority than $j^*$ affected by $C$, we obtain a contradiction to $\hat{\mu}\in\hat{\mathcal{M}}^{j^*}$. 

\item[ \textbf{Case 2}] $C$ makes $j^*$ better off

The cycle $C$ does not affect the welfare of any agent $j<j^*$. Furthermore, since for each agent either all cycles are improving or all cycles are worsening, $C$ cannot violate the CIR constraint of any $j\neq i^*$ at $(A,B)$. Finally, given that there is no blue edge connecting $i^*$ and $o^*$ in $H^*$, $C$ also does not violate the CIR constraint of $i^*$ at $(A,B)$. But then, the matching that we get by implementing all exchanges encoded in $C$ starting from $\mu$ improves $j^*$'s welfare without violating any agent's CIR constraint and we obtain a contradiction to $\mu\in\mathcal{M}^{j^*}$.   
\end{itemize}

For the remainder of the proof, we assume that there is a blue edge
connecting $i^*$ and $o^*$ in $H^*$. Let {$C^*$} be a cycle in $D$ that
affects the welfare of $j^*$. By the
  arguments in Case 1 above, $C^*$ must
  increase the welfare of $j^*$. Furthermore,  $C^*$ has to contain an 
  exchange 
  that involves the blue $i^*o^*$-edge. Note that $C^*$ represents a
  collection of the paths in $H^*$ and may thus involve more agents
  than those incident to the edges of $C^*$ in $D$.

{Since only one cycle in $D$ can contain a path that involves the
  $i^*o^*$ edge,  $C^*$ is the unique cycle that improves $j^*$'s
  welfare.  Because $C^*$ is a cycle, the residual subgraph after
  removing it from $D$, that is $D\setminus C^*$, is also Eulerian can
  again be decomposed into disjoint cycles. Let $\tilde j$ be the
  highest priority agent with positive degree in $D\setminus C^*$ and
  let $\tilde C$ be a cycle that contains $\tilde j$.
  Exactly as argued in Case 1 above for $j^*$, we conclude that
  $\tilde C$ improves $\tilde j$'s welfare. However, 
  since no path in
  $D\setminus C^*$ contains the $i^*o^*$ edge. So $\tilde C$ does not
  violate the CIR constraints of any agent at $(A,B)$, not even
  $i^*$, contradicting $\mu\in \mathcal M^{\tilde j}$. Thus, we deduce
  that $D\setminus C^*$ is empty. In other words, all paths in $H^*$
  appear in a single cycle $C^*$ where each vertex appears only
  once. That is, each agent involved in $C^*$ is an endpoint of
  exactly two paths in $H^*$, red at one and blue at the other. A
  consequence of this is that for each such agent $i$, $|\hat m_i -
  m_i| = 1$ and for each $i$ who does not appear in $C^*$, $|\hat m_i -
  m_i| = 0$. At this point, if $D$ is empty, meaning that the
  introduction of the new edge from an $i^*$ copy to  $o^*$  does not
  affect any agent, then we are done as $|\hat m_i -  m_i| = 0$ for
  every $i$. On the other hand, if there is any $i$ for whom $|\hat
  m_i -  m_i| = 1$, then $C^*$ necessarily involves the $i^*o^*$
  edge, thus establishing that \aw{$o^*\in \hat{\mu}(i^*)$.}

  All that remains to establish are the bounds on $|\hat\mu(i^*)\cap
  \hat A_{i^*}| - |\mu(i^*)\cap  A_{i^*}| $ in the statement of the claim. These bounds depend on
  whether $o^*\in \hat A_{i^*}$ or $o^*\in \hat B_{i^*}$. 
  
  We first consider the case where   $o^*\in \hat A_{i^*}$. If $i^*$
  is not a vertex of $C^*$ in $D$, then $|\hat\mu(i^*)\cap
  \hat A_{i^*}| - |\mu(i^*)\cap  A_{i^*}| = 0$, as claimed. So,
  suppose $i^*$ is a vertex of  $C^*$. If $C^*$ improves $i^*$, then
 $|\hat\mu(i^*)\cap  \hat A_{i^*}| - |\mu(i^*)\cap  A_{i^*}| = 1$,
 which is consistent with the claimed bounds.  For the case of $o^*\in
 \hat A_{i^*}$, all that remains is to show that   $C^*$ cannot worsen
 $i^*$.

 For the sake of contradiction, let us suppose \aw{that $C^*$ worsens
 $i^*$}. Observe that this means that there is slack of at least one in
 $i^*$'s CIR constraint at $\mu$ (for $A_{i^*}$ and $B_{i^*}$)---a
 fact that we will use below. Label the edges in $C^*$  as $e_1, \dots
 , e_M$ , corresponding to paths $P_1, \dots, P_M$ in $H^*$. Assume
 without loss of generality that $e_1 = (i^*, k)$ and\aw{ $e_M = (j,i^*)$}. Since $C^*$ worsens $i^*$,  $\mathbf{blue}^{(i^*,j)}=b(i^*)$ and
 $\mathbf{red}^{(i^*,k)}=a(i^*)$.\footnote{Note that the switch from
   $\mu$ to $\hat \mu$ replaces the first red edge of $e_M$ with the
   last blue edge of $e_1$ meaning that $i^*$ gives up an object in
   $A_{i^*}$ (connected to $a(i^*)$) for one in $B_{i^*}$ (connected
   to $b(i^*)$) .} 
Because we are considering the case where $o^*\in \hat A_{i^*}$, $o^*$
is not incident to the path $P^M$ (because $P^M$, as noted above,
ends at $b(i^*)$ not $a(i^*)$). So, the $i^*o^*$ edge appears in some
other path, $P_{m^*}$ as a blue edge touching $a(i^*)$---it is
therefore not the first edge in $P_{m^*}$. It is not the last edge
either because it touches $a(i^*)$---by our earlier claim since $P_M$
has $i^*$ as a blue endpoint at $b(i^*)$, no path can have  a blue
endpoint at $a(i^*)$ for any other path.  Cut $P_{m^*}$
into two parts at $a(i^*)$: Call the first part, ending with the
$i^*o^*$ edge $P^1_{m^*}$ and call the rest $P^2_{m^*}$.
 Let $\tilde j$ be the highest priority agent affected at the
 interface of any adjacent edges $e_m$ and $e_{m+1}$ for some $m$ such
 that $m^* \leq m < M$. Then $\tilde j$ must be worsened by
 $C^*$. Otherwise, $P^2_{m^*}, P_{m^*+1}, \dots, P_M$ would not
 violate any CIR constraints at $(A,B)$: no agent is worsened any more
 than by $C^*$, which respects CIR constraints and the $i^*o^*$ edge
 is not used. But then we have
 contradicted $\mu\in \mathcal M^{\tilde j}$. Thus, the highest
 priority agent affected by $C^*$ is improved at the interface of
 $P_{m}$ and $P_{m+1}$ for some $m$ such that $1 \leq m < m^*$. But
 then, consider $P_1, \dots, P_{m^*-1}, P^1_{m^*}$: it still improves $j^*$ and
 any other agent improved from $P_1$ to $P_{m^*}$ without worsening
 $i^*$ or $\tilde j$. This contradicts $\hat \mu\in \mathcal
 M^{i^*}$. Therefore, $C^*$ cannot worsen $i^*$ and this completes the
 proof for the case of $o^* \in \hat A_{i^*}$.

 Finally, consider the case where $o^* \in\hat B_{i^*}$. If $i^*$ is
 not a vertex of  $C^*$ there is nothing to prove as the $i^*o^*$ edge
 is on the interior of some path and does not affect the number of
 edges incident to $a(i^*)$ that are selected meaning that
 $|\mu(i^*)\cap A_{i^*}| = |\hat\mu(i^*)\cap A_{i^*}|$. So, suppose
 $i^*$ is a vertex of $C^*$. If $C^*$ worsens $i^*$, there is again
 nothing left to prove since  $|\hat\mu(i^*)\cap A_{i^*}| =
 |\mu(i^*)\cap A_{i^*}|-1$.  We complete the proof by ruling out the
 possibility that  $C^*$ improves $i^*$.Proving that $C^*$
   cannot improve $i^*$ in the case of $o^*\in \hat B_{i^*}$ is very
   similar to the above proof that $C^*$
   cannot worsen $i^*$ in the case of $o^*\in \hat A_{i^*}$. So,
   define $e_1, \dots, e_M$ and $P_1, \dots, P_M$ as in the earlier
   case. Since $C^*$ improves $i^*$, 
   $\mathbf{blue}^{(i^*,j)}=a(i^*)$ and
 $\mathbf{red}^{(i^*,k)}=b(i^*)$. Thus, $o^*\in \hat B_{i^*}$ is not
 incident to $P_1$ and therefore we can, just like we did above,
 define $m^*$   and split it into  $P^1_{m^*}$ and
 $P^2_{m^*}$. The only difference is that the
 terminal blue edge of   $P^1_{m^*}$ is between $b(i^*)$ and $o^*$
and  the terminal red edge of $P^2_{m^*}$ is incident to $b(i^*)$.
Let $\tilde j$ be the highest priority agent affected
at the  interface of some  adjacent edges $e_m$ and $e_{m+1}$ for some
$m$ such  that $m^* \leq m < M$. If $j^*$ is improved, then 
$P^2_{m^*}, P_{m^*+1}, \dots, P_M$ would not
 violate any CIR constraints at $(A,B)$: no agent is worsened any more
 than by $C^*$, which respects CIR constraints and the $i^*o^*$ edge
 is not used. This contradicts 
 $\mu\in \mathcal M^{\tilde j}$. Note that since this short-circuited
 cycle  improves $i^*$ at $(A,B)$, if $\tilde j$ with higher priority
 than $i^*$ is worsened between $P_{m^*}$ and $P_M$, then  we contradict
 $\mu\in \mathcal M^{i^*}$. 
 Thus, we deduce two facts: (1) the highest priority agent affected by
 $C^*$ between $P_{m^*}$ and $P^M$ has higher priority than $i^*$ and is
 worsened; and (2) 
 the highest  priority agent affected by $C^*$, denoted by $j^*$ is
 improved between $P_1$ and $P_{m^*-1}$. But
 then, consider $P_1, \dots, P_{m^*-1}, P^1_{m^*}$: it still improves $j^*$ and
 any other agent improved from $P_1$ to $P_{m^*}$ without worsening
 $\tilde j$. This contradicts $\hat \mu\in \mathcal
 M^{\tilde j}$. Therefore, $C^*$ cannot improve $i^*$ and this
 completes the  proof.
}\end{proof}

We now use the just established claim to complete the proof. Fix an agent $i$ with endowment $\omega(i)$. In the following, we will think of $i$ reporting what is in her top indifference class and all non-endowed objects that are in her second indifference class. That is, if we say that $i$ reports $(A_i,B_i)$ with $A_i\cap B_i=\emptyset$ we actually mean that $i$ reports $(A_i,(\omega(i)\setminus A_i)\cup B_i)$.

In the first part of the proof, consider a triple $(A_i,\hat{A}_i,\hat{B}_i)$ such that $A_i\subseteq \hat{A}_i$ and $\hat{A}_i\cap \hat{B}_i=\emptyset$. We show by induction on $L\equiv|\hat{A}_i\setminus A_i|+|\hat{B}_i|$ that if $i$'s true desirable set is $A_i$, then $(A_i,\emptyset)$ weakly dominates $(\hat{A}_i,\hat{B}_i)$.  Throughout, let $\mu$ be an IRP outcome for the case where $i$ reports $(A_i,\emptyset)$. The base case of $L=1$ follows from the just established claim for $|\hat{B}_i|=1$ and the results on strongly trichotomous preferences in \cite{ManjunathWestkamp:2018a} for $|\hat{A}_i|=1$. So suppose $L\geq 2$ and that the statement has already been established for all manipulations that add fewer than $L$ objects to the top-two indifference classes. Let $(\hat{A}_i,\hat{B}_i)$ be such that $\hat{A}_i\cup\hat{B}_i=\{o^1,\dots,o^{L-1}\}$, $\hat{\mu}$ be an IRP outcome for when $i$ reports $(\hat{A}_i,\hat{B}_i)$ and let $o^L\in O\setminus (\hat{A}_i\cup\hat{B}_i)$ be arbitrary. First, consider the report $(\hat{A}_i,\hat{B}_i\cup\{o^L\})$ and let $\hat{\mu}'$ be a corresponding IRP outcome. By the claim established above, we have that $|\hat{\mu}(i)\cap (A_i\cup \hat{A}_i)|\geq |\hat{\mu}'(i)\cap (A_i\cup \hat{A}_i)|$. Hence, if contrary to what we aim to show, $|\hat{\mu}'(i)\cap A_i|>|\mu(i)\cap A_i|$, there has to be some object $p\in (\hat{A}_i\cap \hat{\mu}(i))\setminus (A_i\cup \hat{\mu}'(i))$ and some object $q\in (A_i\cap \hat{\mu}'(i))\setminus \hat{\mu}(i)$.  By the rules of IRP, $\hat{\mu}'$ is still an IRP outcome for the report $(\hat{A}_i\setminus\{p\},\hat{B}_i\cup\{o^L\})$. Since $|\hat{A}_i\setminus (A_i\cup \{p\})|+|\hat{B}_i\cup\{o^L\}|=L-1$, we obtain a contradiction to our inductive assumption. Second, consider the report $(\hat{A}_i\cup\{o^L\},\hat{B}_i)$ and let $\hat{\mu}'$ again denote an IRP outcome for this report. Assume again that $|\hat{\mu}'(i)\cap A_i|>|\mu(i)\cap A_i|$.  By the inductive assumption, $|\mu(i)\cap A_i|\geq |\hat{\mu}(i)\cap A_i|$. Furthermore, extrapolating from our arguments in the previous case, we must have $\hat{A}_i\subseteq \hat{\mu}'(i)$: Otherwise, we would again obtain a contradiction to the inductive assumption when we delete an object from $\hat{A}_i\setminus \hat{\mu}'(i)$. Now consider the report $(A_i\cup \hat{A}_i,\omega(i)\setminus (A_i\cup\hat{A}_i))$ and let $\tilde{\mu}$ be a corresponding IRP outcome. By the inductive assumption, we have that $|\mu(i)\cap A_i|\geq |\tilde{\mu}(i)\cap A_i|$. Hence, $|\hat{\mu}'(i)\cap (A_i\cup \hat{A}_i)|>|\tilde{\mu}(i)\cap (A_i\cup \hat{A}_i)|$ and we obtain a contradiction to the inductive assumption for the triple $(A_i\cup \hat{A}_i,A_i\cup\hat{A}_i\cup\{o^L\},\hat{B}_i)$ if $\hat{A}_i\neq \emptyset$. If $\hat{A}_i=\emptyset$, we obtain a contradiction to the claim established above since we must have $\hat{B}_i=\{o^1,\dots,o^{L-1}\}$ and the inductive assumption implies that reporting $(A_i\cup \{o^L\},\{o^1,\dots,o^{L-2}\})$ is worse for $i$ than $(A_i,\emptyset)$ while - as per our assumption - $(A_i\cup\{o^L\},\hat{B}_i)$ is strictly better for $i$.  

Next, we consider arbitrary triples $(A_i,\hat{A}_i,\hat{B}_i)$ such that $\hat{A}_i\cap\hat{B}_i=\emptyset$. By what we have just shown, it is without loss of generality to assume $\hat{A}_i\subseteq A_i$ (if not, consider $\tilde{A}_i\equiv A_i\cap \hat{A}_i$ and note that $(\tilde{A}_i,\emptyset)$ weakly dominates $(\hat{A}_i,\hat{B}_i)$ by what we have already established). By the results for the strongly trichotomous case in \cite{ManjunathWestkamp:2018a}, $(A_i,\emptyset)$ weakly dominates $(\hat{A}_i,\emptyset)$. By the above, $(\hat{A}_i,\emptyset)$ weakly dominates $(\hat{A}_i,\hat{B}_i)$. Together, these last two statements complete the proof.

\end{document}